\documentclass[useAMS,usegraphicx]{mn2e}
\pdfoutput=1                  %MY ADDITION --> forces PDFTex (for arxiv)
\usepackage{rotate}
\usepackage{rotating}
\usepackage{times}
\usepackage{cite}             %MY ADDITION --> for BibTex
\usepackage{comment}  %MY ADDITION --> allows bulk commenting
\usepackage{graphicx}    %MY ADDITION --> allows cropping of figures
\usepackage{color}          %MY ADDITION --> allows colored text
\usepackage{bm}            %MY ADDITION --> allows bold text in math mode
\newif\ifAMStwofonts
\AMStwofontstrue
\def\mrk493{{Mrk~493}}

\def\kdblur2{\textsc{kdblur2}}

\def\kdblur{\textsc{kdblur2}}

\def\relline{\textsc{relline}}
\def\xillver{\textsc{xillver}}
\def\relxill{\textsc{relxill}}

\def\Msun{M$_{\odot}$}
\def\Mbh{\thinspace \emph{M$_{\rm{BH}}$}}

\def\suzaku{{\it Suzaku }}

\def\xmm{{\it XMM-Newton }}
\def\astroh{{\it Astro-H }}          %added
\def\nustar{{\it NuStar }}           %added
\def\redchisq{{$\chi^2_\nu$}}

\def\feka{{Fe~K$\alpha$}}

\def\Rg{{$\rm\thinspace R_{g}$}}     %added
\def\deg{$^{\circ}$}                 %alternative syntax for this above when I made it

\def\cm{{\rm\thinspace cm}}
\def\erg{{\rm\thinspace erg}}
\def\eV{{\rm\thinspace eV}}

\def\ga{{\rm\thinspace gauss}}

\def\keV{{\rm\thinspace keV}}

\def\Msun{{\hbox{$\rm\thinspace M_{\odot}$}}} %edited for error

\def\s{{\rm\thinspace s}}
\def\ergcmps{\hbox{$\erg\cm\s^{-1}\,$}}                        %added 

\title{How well can we measure supermassive black hole spin?}

\author[K. Bonson \& L. C. Gallo]{K. Bonson$^1$ and L. C. Gallo$^1$  
               \\ 
$^{1}$ Department of Astronomy and Physics, Saint Mary's University, 923 Robie Street, Halifax, NS, B3H 3C3, Canada \\
}
\pagerange{\pageref{firstpage}--\pageref{lastpage}}
\begin{document}
\maketitle
\label{firstpage}

\begin{abstract}
Being one of only two fundamental properties black holes possess, the spin of supermassive black holes (SMBHs) is of great interest for understanding accretion processes and galaxy evolution. However, in these early days of spin measurements, consistency and reproducibility of spin constraints have been a challenge. Here we focus on X-ray spectral modeling of active galactic nuclei (AGN), examining how well we can truly return known reflection parameters such as spin under standard conditions. We have created and fit over 4000 simulated Seyfert 1 spectra each with 375$\pm$1k counts. We assess the fits with reflection fraction of $R$ = 1 as well as reflection-dominated AGN with $R$ = 5. We also examine the consequence of permitting fits to search for retrograde spin. In general, we discover that most parameters are over-estimated when spectroscopy is restricted to the 2.5 -- 10.0\keV\  regime and that models are insensitive to inner emissivity index and ionization. When the bandpass is extended out to 70\keV, parameters are more accurately estimated. Repeating the process for $R$ = 5 reduces our ability to measure photon index ($\sim$3 to 8 per cent error and overestimated), but increases precision in all other parameters --- most notably ionization, which becomes better constrained ($\pm$45\ergcmps) for low ionization parameters ($\xi$$<$200\ergcmps). In all cases, we find the spin parameter is only well measured for the most rapidly rotating supermassive black holes (i.e. $a$ $>$ 0.8 to about $\pm$0.10) and that inner emissivity index is never well constrained. Allowing our model to search for retrograde spin did not improve the results.

\end{abstract}


%------------------------------------------------------------------------------------
%------------------------------------------------------------------------------------

\section{Introduction}
\label{sect:intro}
It is believed that most black holes will be ``born'' with some amount of angular momentum, instilled in them from their progenitors (Kerr 1963). This angular momentum can change over time, spinning-up the black hole through prograde accretion of matter or spinning-down through mergers (e.g. Volonteri et al. 2013; King \& Pringle 2006, 2007; King 2008). Black hole spin, defined by the dimensionless spin parameter: $a = Jc/GM^2$ with theoretical values ranging -0.998 $<$ a $<$ 0.998, is a parameter of extreme interest. The classical Thorne limit quoted here (Thorne 1974) does not include modern magnetohydrodynamic (MHD) accretion theory. If MHD is considered, the limit reduces to $a\sim$0.95 (e.g. Reynolds et al. 2006). For the last decade or so, sophisticated spectral models and high quality data make it possible to measure the black hole spin parameter in active galactic nuclei (hereafter AGN; e.g. Brenneman \& Reynolds 2006).

%%%%%%%%%%%%%%% Table: input xillver model %%%%%%%%%
\begin{table*}
\setlength{\tabcolsep}{12pt}
\centering{
\caption{Model details for simulated spectral analysis. Parameters that are permitted to vary are in bold. Simulated spectra were created with all key parameters generated randomly within the given ranges. The simulated spectra were then fit with a model whose default parameters were based off of those for average Seyfert 1 AGN. The key parameters were allowed to vary during the fitting process, while both $q_{1}$ and $\xi$ were fixed for different versions of fit tests. Parameter values that do not change from the initial Test A are denoted with dashes. All four tests were performed for reflection fractions of $R$ = 1 and $R$ = 5.}
\scalebox{1.0}{       
	\begin{tabular}{ccccccc}
	\hline
	\bf{Parameter} & \bf{Input \textbf{Range}} & \multicolumn{4}{c}{\bf{Fit Default}} & \bf{Units} \\
	 && \bf{Test A} & \bf{Test B} & \bf{Test C} & \bf{Test D} & \\ \hline
	 \bf{inner emissivity ($\bf{q_{1}}$)} & \bf{3 - 9} & \bf{3.0} & --- & 3.0 & 3.0 &     \\
	 outer emissivity (\emph{$q_{\rm{2}}$})  & 3.2 & 3.0 & --- & --- & --- & \\
	 break radius (\emph{$R_{\rm{br}}$}) & 4.8 & 6.0 & --- & --- & --- & $R_{\rm{g}}$ = $\frac{GM}{c^{2}}$ \\
	 \bf{BH spin (\emph{a})} & \bf{0 - 0.998} & \bf{0.5} & --- & --- & --- &\\
	 \bf{inclination (\emph{$\theta$})} & \bf{20 - 70} & \bf{30} & --- & --- & --- & deg \\
	 inner disk radius (\emph{$R_{\rm{in}}$}) & 1 & 1&  --- & --- & --- & \emph{$R_{\rm{ISCO}}$} \\
	 outer disk radius (\emph{$R_{\rm{out}}$}) & 400 & 400 & --- & --- & --- & $R_{\rm{g}}$ \\
	 redshift (\emph{z}) & 0.05 & 0.05 & --- & --- & --- &\\
	 \bf{photon index (\emph{$\Gamma$})} & \bf{1.7 - 2.2} & \bf{2.0} & --- & --- & --- & \\
	 \bf{ionization (\emph{$\xi$})} & \bf{50 - 500} & \bf{75} & 75 & --- & 75 & erg cm $s^{-1}$ \\
	 \bf{iron abundance ($A_{Fe}$)} & \bf{0.5 - 5.0} & \bf{3.0} & --- & --- & --- & solar \\
	 \label{models}
	 \end{tabular}}}
\end{table*}
%%%%%%%%%%%%%%%%%%%%%%%%%%%%%%%%%%%%%

Supermassive (\Mbh $\geq10^{6}$\Msun) black hole (hereafter SMBH) spin in particular may have powerful implications on a wide range of scales, from close to the black hole itself out to the host galaxy due to its direct influence on how mass is accreted in these objects (e.g. Cappi 2006, Davis \& Laor 2011, Gabor \& Bournaud 2014, Bourne et al. 2014). Studies have shown that accretion flow can significantly affect mass ejection from the central engine of AGN, potentially in the form of high-velocity ($\sim0.1c$) winds (Gofford et al. 2015) and/or radio jets (Blandford \& Znajek 1977, Turner \& Shabala 2015, King et al. 2015). 

These different forms of mechanical AGN feedback, along with intense radiation emitted from the central engine, appear to influence star formation in the host galaxy by means of galaxy self-regulation (Martizzi et al. 2013, Taylor \& Kobayashi 2015) and can be observed in host-black hole virial relations such as the M-$\sigma$ relation (e.g. Gebhardt et al. 2000) and black hole fundamental plane (e.g. Merloni et al. 2003) and may provide the key to SMBH-host galaxy co-evolution. In addition to the cosmological implications, environmental conditions close to a Kerr black hole are some of the most extreme in the Universe. They provide us with the unique opportunity to analyze more exotic physical phenomena such as light-bending (e.g. Miniutti \& Fabian 2004, Wilkins \& Fabian 2012, Gallo et al. 2013) and reverberation delays (e.g. Fabian et al. 2009, Zoghbi et al. 2010). 

Although the typical spin parameter and the fraction of spinning SMBH are still unknown, the vast majority of current measurements from SMBHs have a high ($a > 0.8$) spin (Brenneman 2013, Reynolds 2014, Vasudevan et al. 2015). However, large-scale survey analyses are limited due to sampling bias and many questions remain as to the true distribution of AGN spin.

Spin measurements are becoming more commonplace as the number of quality spectra from AGN continues to grow. While black hole spin can, in theory, be constrained in a variety of ways such as continuum fitting (e.g. Done et al. 2013), the broad \feka\ line (e.g. Walton et al. 2013, Gallo et al. 2015), and potentially quasi-periodic oscillations or QPOs in stellar-mass black holes (e.g. Mohan \& Mangalam 2014), our most robust measurements to date rely on our ability to detect a strong reflection component in the X-ray spectra. The \feka\ line, at 6.4\keV\ in the source rest-frame, can act as a probe of the innermost regions of the AGN accretion disk: its profile containing information on disk ionization and abundances (e.g. Reynolds et al. 2012, Bonson et al. 2015), inclination and reflection strengths (e.g. Walton et al 2013), and disk emissivity (Wilkins \& Fabian 2012, Wilkins et al. 2014). 

As with any technique, there are assumptions that go into constraining spin using the \feka\ line. It is assumed that the emission we observe from the broadest component of the line is coming from the innermost stable circular orbit (ISCO) and that there is a negligible radiative contribution from within the ISCO. The accretion disk is considered to be the standard ``Shukura-Sunyaev disk'' -- i.e. ideally thin, ionized, and isothermal (Shukura \& Sunyaev 1973) -- and that gravitational forces from the central black hole dominates. These assumptions seem reasonable for all but the most extreme scenarios and do well to model what is in reality a very complex region.

The X-ray instruments on-board \xmm and \suzaku  are ideal for constraining spin in the manner described above because of their superior sensitivity in the 2 to 10\keV\ band. Indeed, most SMBH spin measurements in the literature today utilize \xmm and \suzaku data. Now, with \nustar and \astroh extending observations into the hard X-ray regime up to 80\keV, even more of the reflection spectrum can be resolved and analyzed for more accurate modelling and, thus, spin constraints. However, as measurements are repeated, we find in some cases inconsistent spin measurements for a given AGN. For example, in the case of MCG-06-30-15, this broad-line Seyfert 1 galaxy has been studied thoroughly and its spin has been measured multiple times (e.g. Walton et al. 2013, Patrick et al. 2012, Brenneman \& Reynolds 2006). However, spin measurements of MCG-06-30-15 have varied from an extreme limit of $a>0.98$ to being low-to-moderate at $0.49^{+0.20}_{-0.10}$. Spin analysis for the broad-line radio galaxy 3C~120 is even more contradictory: a prograde spin ($a$ = 0.95) being just as likely as a retrograde ($a$ $<$ 0.10) in the same study (Cowperthwaite \& Reynolds 2012; see Lohfink et al. 2013 for alternative arguments for prograde spin). 

There are known difficulties in modelling AGN spectra and constraining spin. For example, disk ionization, iron abundance, and reflection fraction can all influence the strength of the \feka\ line compared to the continuum. The contrast between the line and the continuum will decrease with increasing spin as general relativistic effects begin to dominate, broadening and redshifting an intrinsically narrow feature. Including further intrinsic spectral complexities like partial covering absorbers, outflows, and distant reflection to the already-challenging fits process and it is easy to see why it can be difficult to constrain spin with even the highest-quality data. In addition, we have no standardized procedure for spin measurements using the \feka\ line -- understandable considering the variation and complexity exhibited in the range of objects we observe. There have been several reviews published over the years, which provide some guidance on how best to approach measuring spin (e.g. Brenneman 2013, Reynolds 2013). Obviously good data are required. For example, Guainazzi et al. (2006) found that an observation of 200k counts or higher in the 2 -- 10\keV\ band and a broad line equivalent width of at least 100\eV\ were required for robust detection of a relativistically-blurred \feka\ feature. These predictions appear to be supported by the current literature.

In this work, we test how reliably we can measure spin and other spectral parameters, themselves important in constraining $a$. We test the influence of bandpass in our measurements, specifically looking at the Compton hump regime and its effect on modelling the reflection component. The key questions we will be asking ourselves include: Under which conditions can we be the most confident in our parameter fits, spin or otherwise? Which energy bands are most conducive to model fitting? Are there any steps we can take to limit parameter degeneracies?

This paper is organized as follows: Section \ref{sim_details} describes how the simulated spectral data were autonomously produced and fit, including a detailed review of the different analysis tests performed for reflection fraction, bandpass, and retrograde spin. Section \ref{R1} provides a step-by-step description of the test results for a reflection fraction of $R$ = 1, including both 2.5 -- 10\keV\ and 2.5 -- 70\keV\ spectral fitting, and Section \ref{R5} repeats the process for the $R$ = 5 scenario. The results of our retrograde spin tests are discussed in Section \ref{retrograde}. Section \ref{discussion} discusses the implications of this work, including caveats and limitations, and conclusions are stated in Section \ref{conclusion} along with future work. 

%------------------------------------------------------------------------------------

\section{\textbf{Simulations}}
\label{sim_details}
Ideally one would use a control with known parameters in order to examine the accuracy of a computational model. Unfortunately, one cannot place an AGN in a laboratory and determine its intrinsic properties to use as a baseline. It is possible, however, to simulate a simple X-ray spectrum having expected characteristics of an average AGN and then fit the simulated spectrum using common techniques in order to examine the reproducibility of model parameters such as spin.

We simulated AGN spectra in the 0.01 -- 300.0\keV\ band with \xmm pn response using the model \relxill: a combination of the reflection model \xillver\ (Garcia et al. 2013) and the \relline\ code (Dauser et al. 2013a) for relativistic blurring. In order to ensure that data quality and signal-to-noise were not limiting factors, analysis was performed on high quality spectra ($350,000\pm1,000$ counts in both the 2.5 -- 10\keV\ and 10 -- 70\keV\ bands) in order to mimic the best observational data currently in hand. At this stage, we did not include more complicated model components such as Galactic absorption, warm absorbers, partial covering absorbers, or distant reflectors. Galactic absorption would influence the SED below 1\keV, a regime that is not addressed at this time, and additional reflectors or absorbers -- as common as they are empirically -- would only serve to complicate the simulated spectra further. We must begin by assessing the performance of AGN spectral modelling in the simplest of scenarios to be most constructive.

The following key parameters were varied during the creation of the spectral simulations: photon index ($\Gamma$, where $N(E) \propto E^{-\Gamma}$ is the incident flux), inner emissivity index ($q_{1}$), black hole spin ($a$), disk inclination angle ($\theta$), ionization ($\xi=4\pi F/n$ where $F$ is flux and $n$ is the hydrogen number density), and iron abundance ($A_{Fe}$) in solar units (see Table\thinspace\ref{models} for details). Varying all six parameters at once, a random number generator produced values within a given range for each parameter for a specified number of spectra. Running error calculations on each key parameter is time consuming for the number of spectra analyzed and we rely on the sampling statistics to reasonably represent the random error in the fitting parameters. The influence of individual error checks and the effect of local minima are discussed further in Section \ref{caveats}.

Once produced, the simulated spectra were then fit with the \xillver\ model. Key parameters set at default starting values before allowing to vary in a step-wise fashion mimicking manual fitting procedures. The potential scope of this study is nearly limitless. To focus the analysis and provide ourselves with a baseline from which to expand work in the future, we considered four primary model fitting tests: Test A allowed all six key parameters to vary, Test B kept $\xi$ fixed at 75\ergcmps, Test C kept $q_{in}$ fixed at 3, and Test D kept both $\xi$ and $q_{in}$ fixed at the aforementioned values (see Table \ref{models}).

We also considered what effect an extended spectral band would have by utilizing \nustar response matrices for the FPMA and FPMB detectors (Harrison et al. 2013). In the reflection scenario, the Compton hump illustrates the balance between photon scattering and absorption in the accretion disk: at energies below $\sim$10\keV, any scattered light is absorbed by metals in the disk. Photoelectric absorption diminishes above 10\keV\ and scattering dominates, appearing in the reflected spectrum as a hump which peaks between 30 -- 40\keV. Around 40\keV, the reflected spectrum turns over due to Compton recoil. By extending our analysis into the Compton Hump regime, we are providing more information on the reflected spectrum and thus should be able to better constrain reflection parameters.

The simulated spectra that were initially fit between 2.5 -- 10\keV\ were then refit between 2.5 -- 70\keV. It should be emphasized that this does not fully simulate an actual XMM-\nustar simultaneous observational analysis, nor was that the intent; we wanted to examine how inclusion of the Compton hump feature influenced our ability to constrain black hole spin. This continues to keep the analysis as instrument-independent as possible and allows us to remain conservative in our approach. In effect, the data were not allowed to overlap. The \emph{XMM}-made spectra were used for E $<$ 10\keV, the \emph{NuStar}-made data were used for E $>$ 10\keV. Spectra were normalized to the same flux and cross-calibration effects were not considered.

In addition, the influence of the reflection fraction ($R$) was examined. Reflection fraction is defined as the ratio of reflected flux over primary flux (Dauser et al. 2013). The blurred reflection interpretation of AGN spectra suggests that our ability to measure disk parameters -- spin, inclination, ionization, and emissivity -- should improve as reflection fraction increases. As more of the total X-ray flux is emitted by reflection off the accretion disk, dominating the illuminating power law continuum, key features such as the \feka\ line will be stronger and more of the overall broadband reflection profile is available to be modelled. In order to test this theory, the procedure outlined above for the $R$ = 1 scenario was repeated for simulated spectra with $R$ = 5. Reflection fractions as high as 10 have been reported in the literature (e.g. Mrk~335, Gallo et al. 2015) and it is not unusual for more exotic AGN, such as narrow-line Seyferts, to present $R$-values larger than 5. Thus, in an effort to remain consistent with our ``average'' Seyfert galaxy spectra and still simulate significantly higher reflection fractions, a value of $R$ = 5 appears to be a reasonable choice.

Lastly, we expect measurements to be sensitive to the possibility of retrograde spin and so some basic retrograde spin analyses are performed for each reflection fraction. Empirical studies have suggested that it is reasonable to suspect that the vast majority of SMBHs have prograde spin, due to the spin-up affect of accretion. That said, there can indeed be cases where a retrograde spinning black hole could be found -- such as the immediate aftermath of a SMBH binary merging event (e.g. Hughes \& Blandford 2003, Miller \& Krolik 2013). In the case of a retrograde spin, the ISCO recedes from the event horizon and can be found at $\sim$9\Rg\ or further. Therefore, there is some justification for allowing the fit model to extend into the retrograde regime and see how that may, or may not, influence our ability to analyze these data overall.

In summary, we created four fit tests (A, B, C, and D) in the 2.5 -- 10\keV\ regime, which we ran for cases of $R$ = 1 and $R$ = 5. We repeated this procedure for an extended energy band of 2.5 -- 70\keV\ to explore the importance of the Compton Hump. We lastly also repeated the broadband fits to examine the influence of allowing the fits to search for retrograde spin.

%%%%%%%%%% Figure: XMM only R1TA Combo Scatter + Bin Plots %%%%%%%%%
\begin{figure*}
	\centering
	%\minipage{1.0\pagewidth}
		\vspace{-1in}
		\advance\leftskip-3.5in   
		\begin{minipage}{0.5\linewidth}
			\includegraphics[scale=0.45]{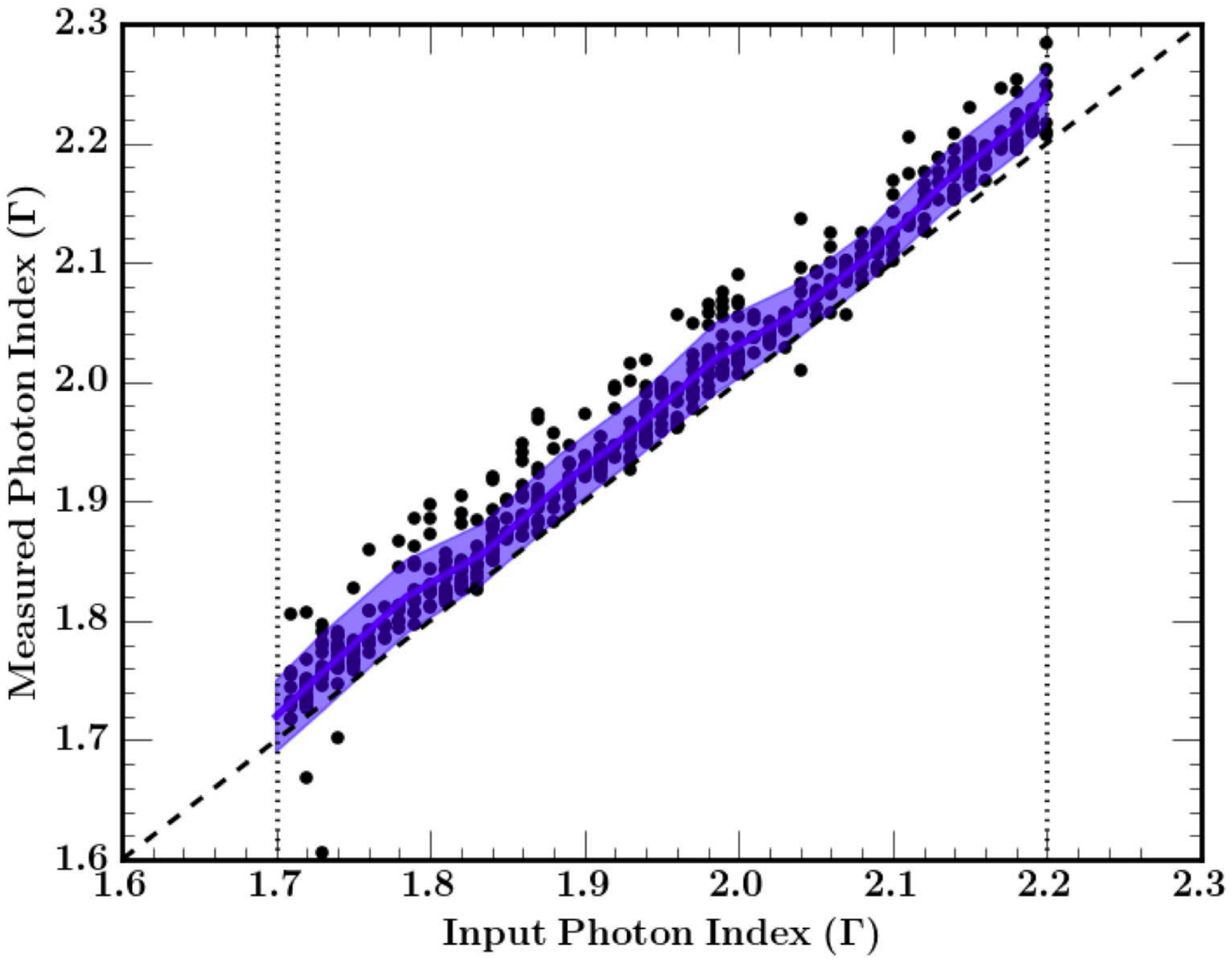}
		\end{minipage} 
		\begin{minipage}{0.01\linewidth}
			\includegraphics[scale=0.45]{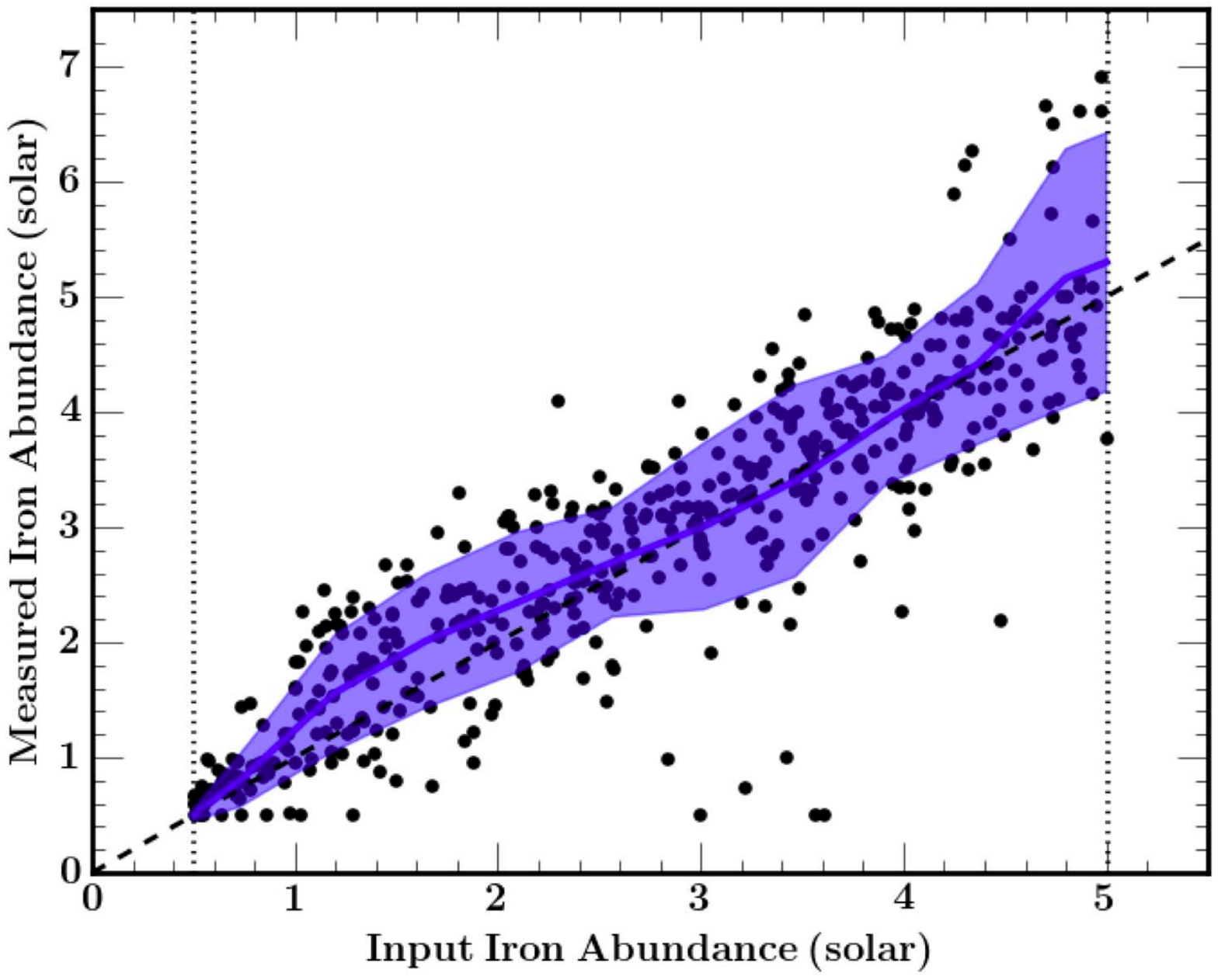}
		\end{minipage} \\
		\vspace{-2.25in}
		\begin{minipage}{0.5\linewidth}
			\includegraphics[scale=0.45]{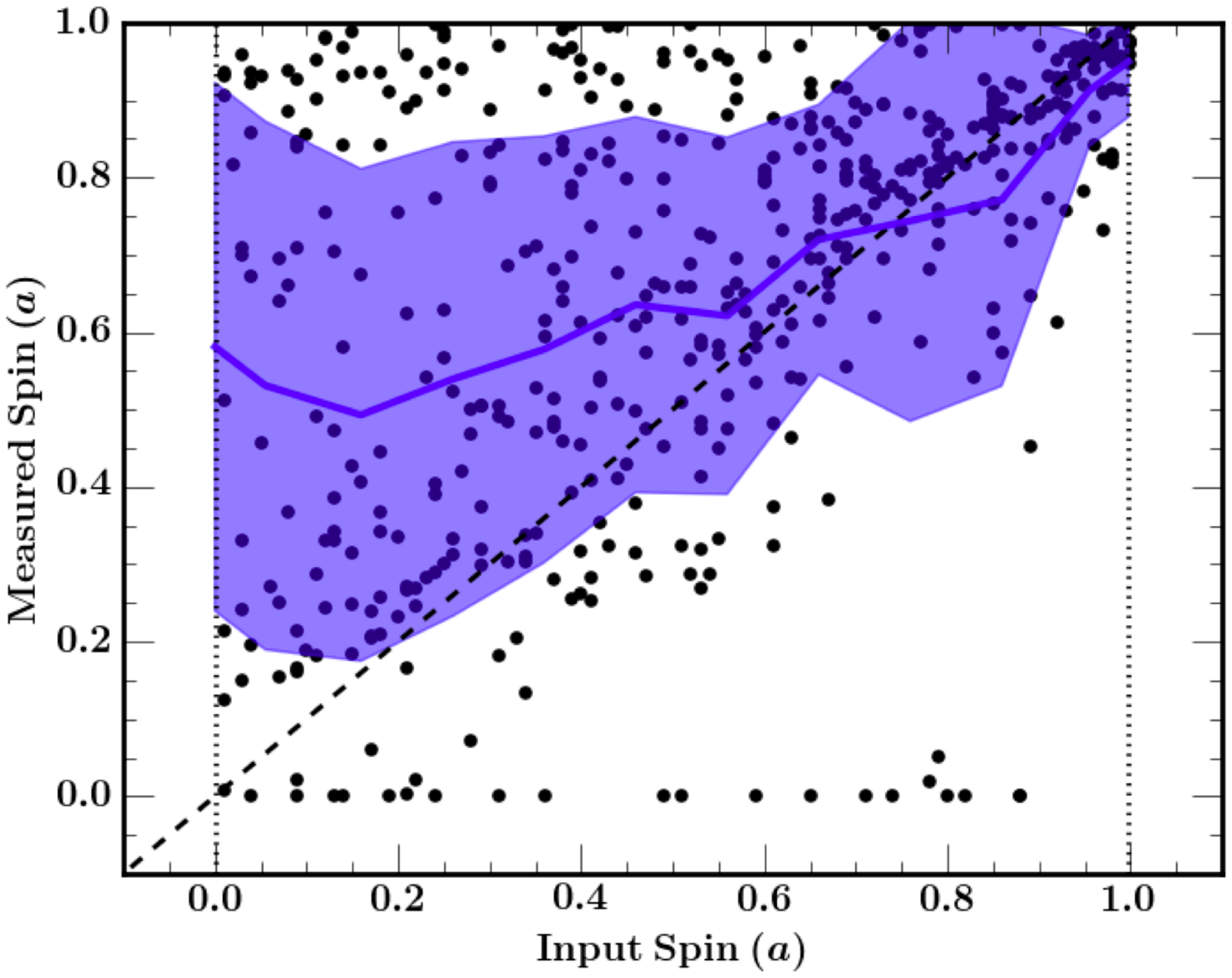}
		\end{minipage}  
		\begin{minipage}{0.01\linewidth}
			\includegraphics[scale=0.45]{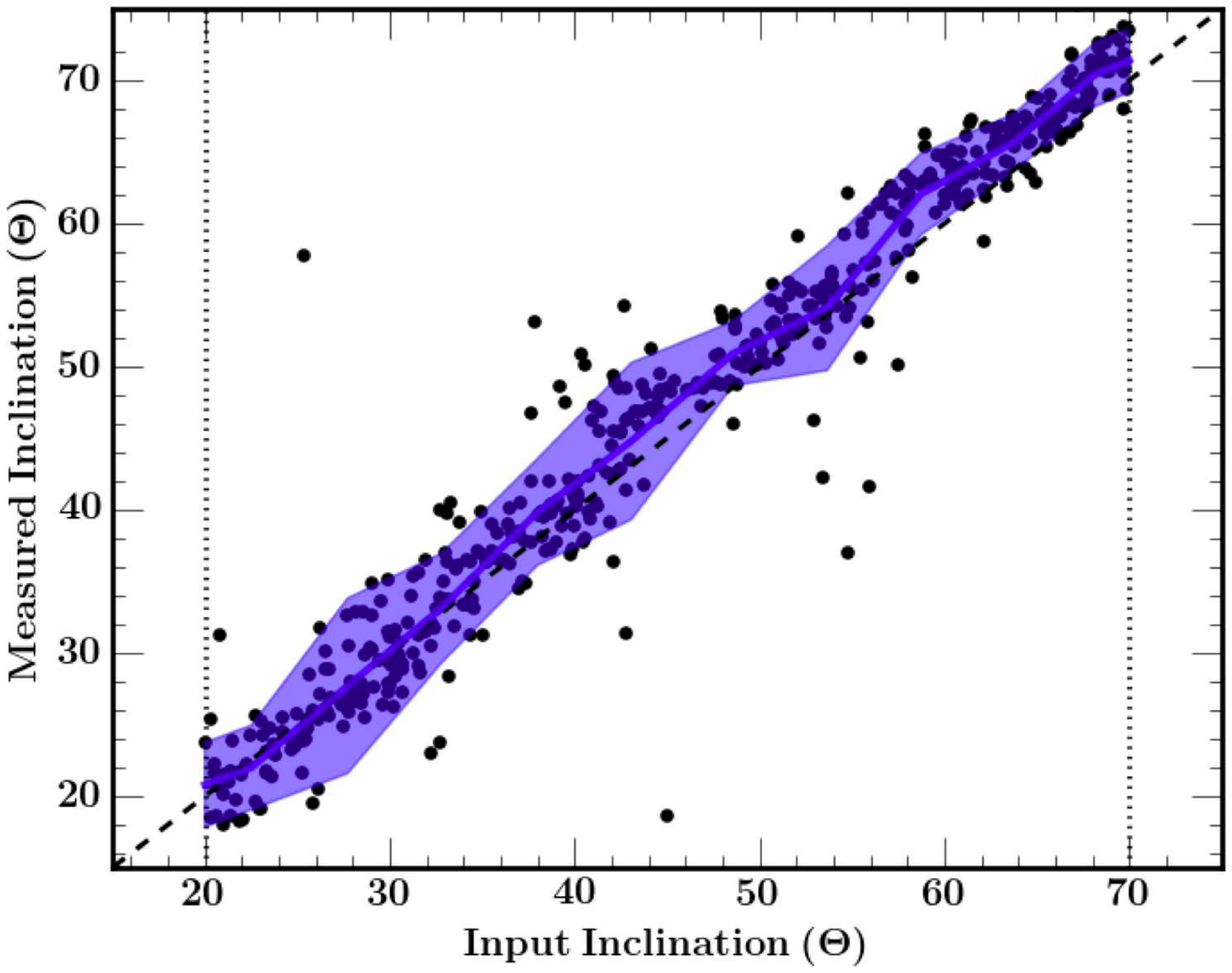}
		\end{minipage}\\
		\vspace{-2.25in}
		\begin{minipage}{0.5\linewidth}
			\includegraphics[scale=0.45]{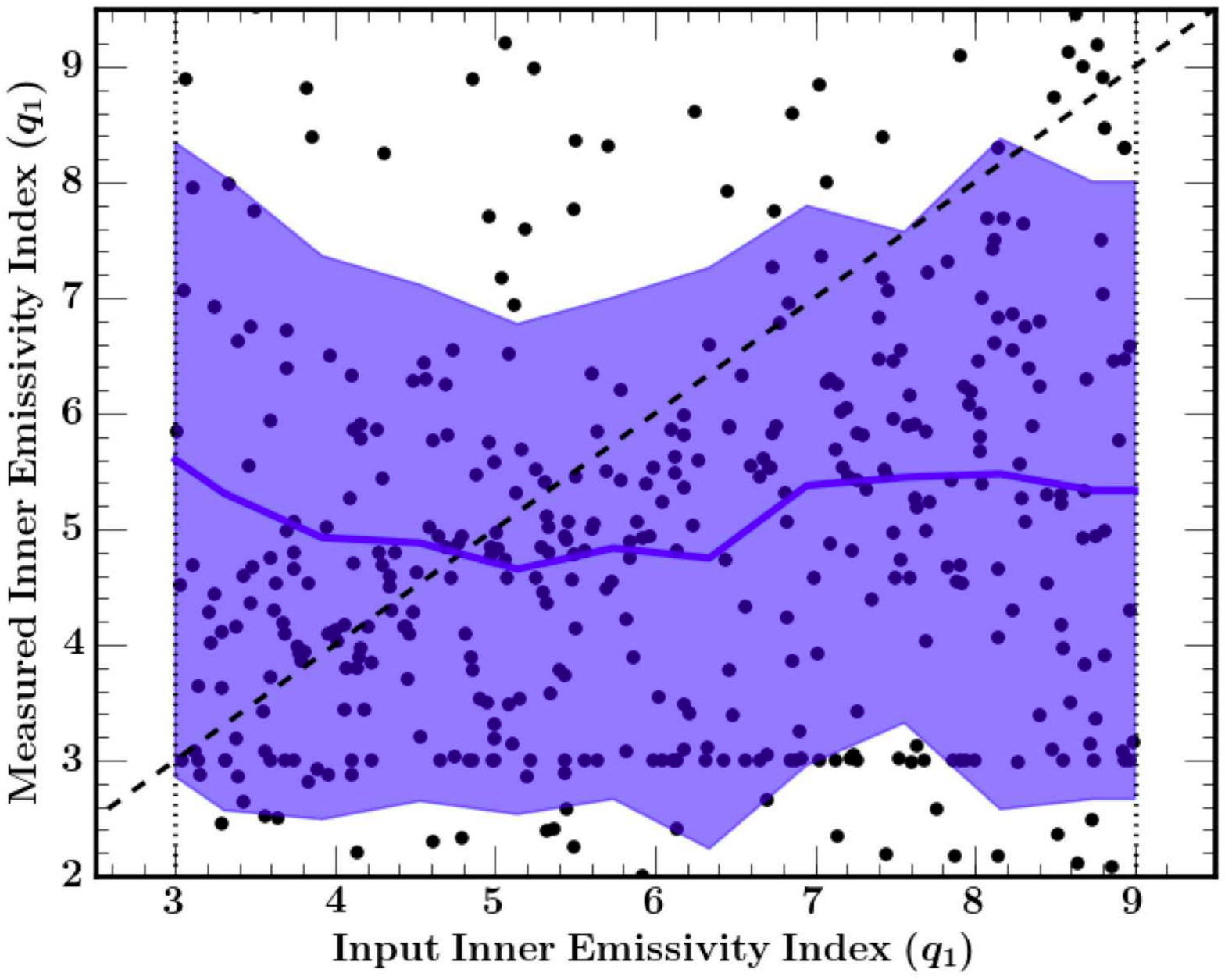}
		\end{minipage}  
		\begin{minipage}{0.01\linewidth}
			\includegraphics[scale=0.45]{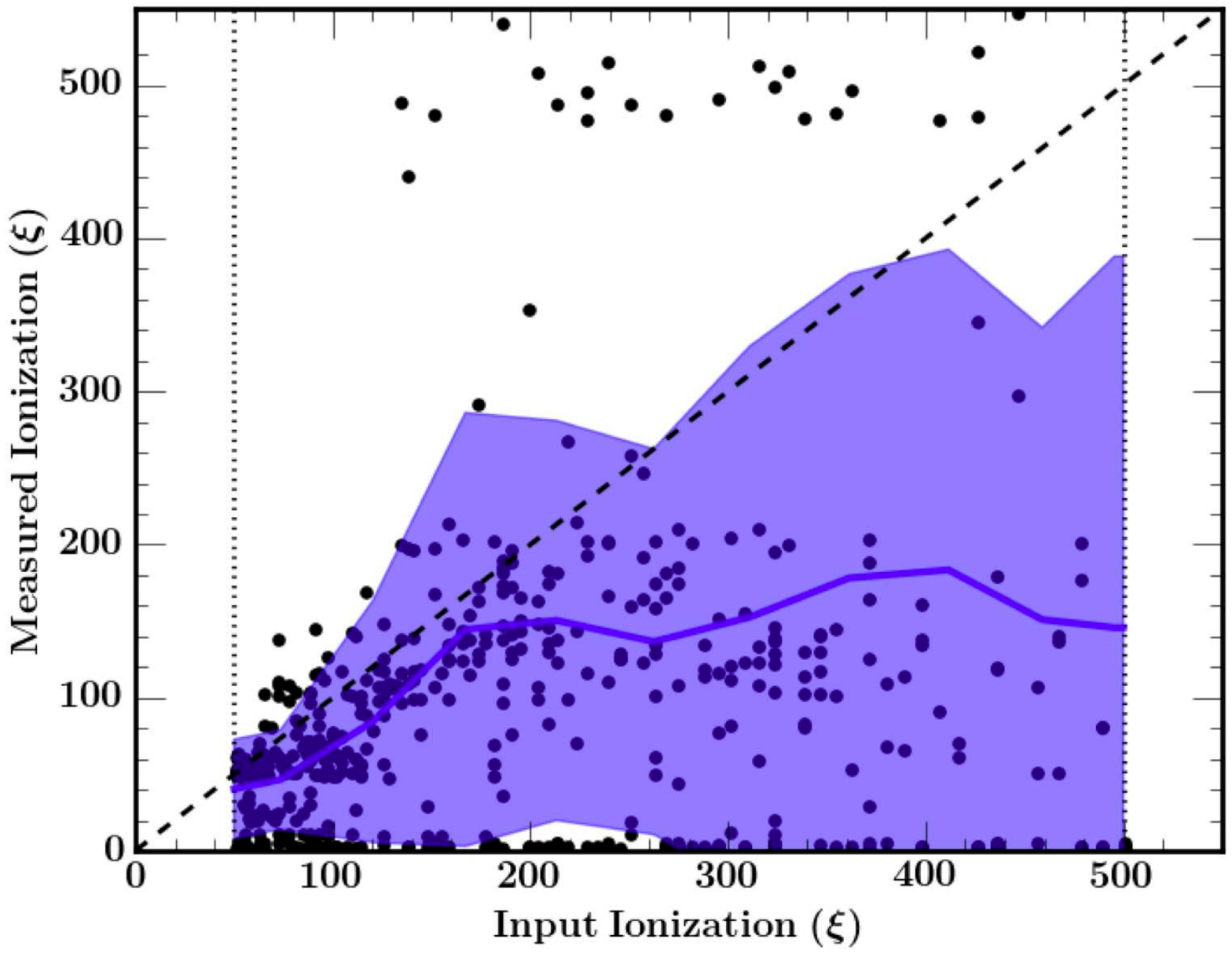}
		\end{minipage} 
		\vspace{-1in}
	%\end{minipage}
    	\caption{The results of the $R$ = 1 simulated spectral fitting from Test A, where all 6 parameters were free to vary. Plots show input parameter values on the abscissa and the measured values on the ordinate. The dashed 1:1 line represents a perfect measurement and the dotted lines denote allowed input ranges. Simulated spectra were produced using the \xmm pn response and fit from 2.5 -- 10\keV\ using the XSPEC model \textsc{relxill} for a collection of randomly-generated input parameters. Each data point corresponds to a modelled spectrum with \redchisq $<$ 1.1. The data were binned by input value and overplotted as blue bands, the widths of which represent 1$\sigma$ error.}
   	 \label{R1XMMcombo}
\end{figure*}
%%%%%%%%%%%%%%%%%%%%%%%%%%%%%%%%%%

%%%%%%%%%% Figure: XMM only R1TA binned plots %%%%%%%%%
\begin{figure*}
	\centering
	\vspace{-1in}
	\advance\leftskip-3.5in   
	\begin{minipage}{0.5\linewidth}
		\scalebox{0.45}{\includegraphics{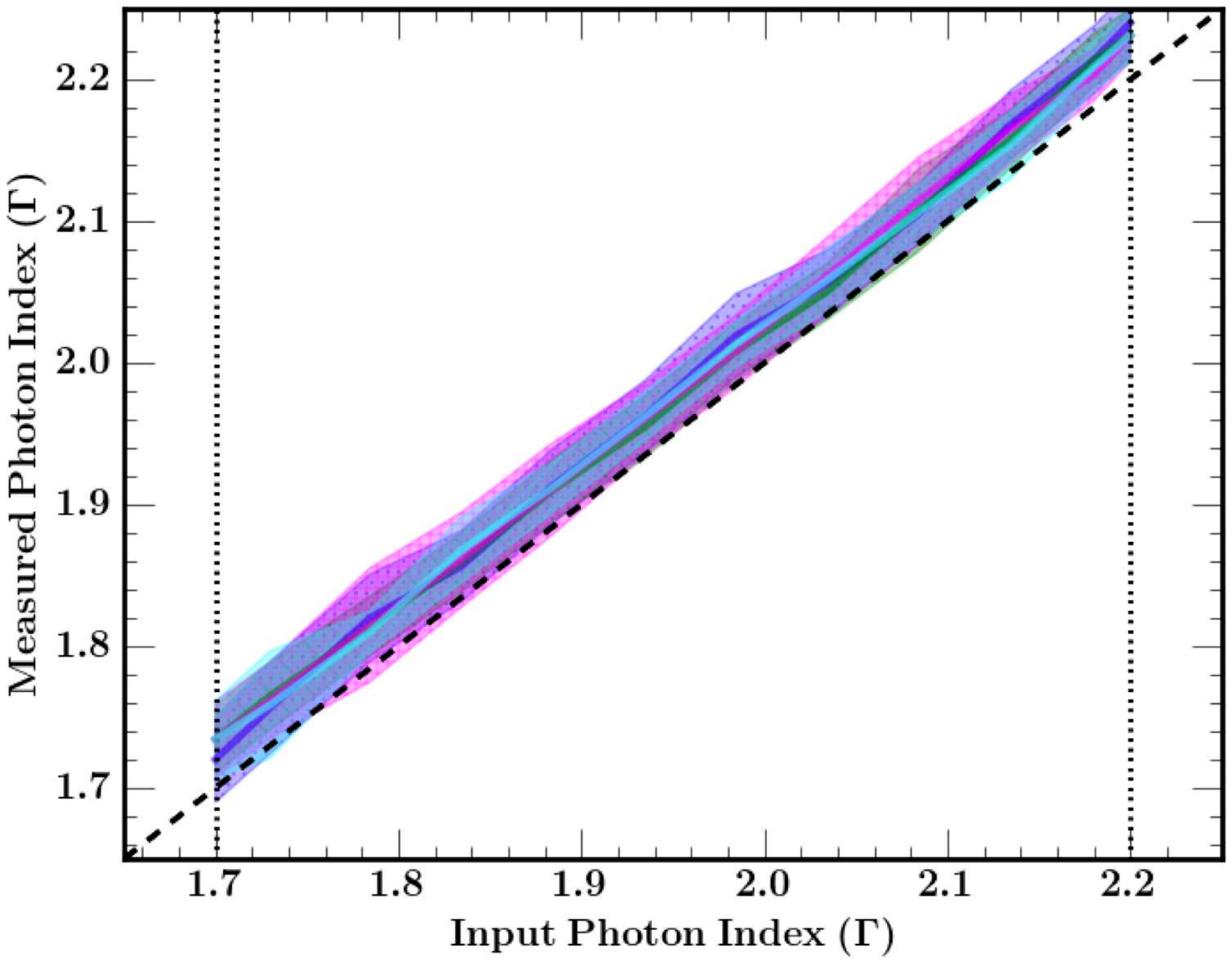}}
	\end{minipage}  
	\begin{minipage}{0.01\linewidth}
		\scalebox{0.45}{\includegraphics{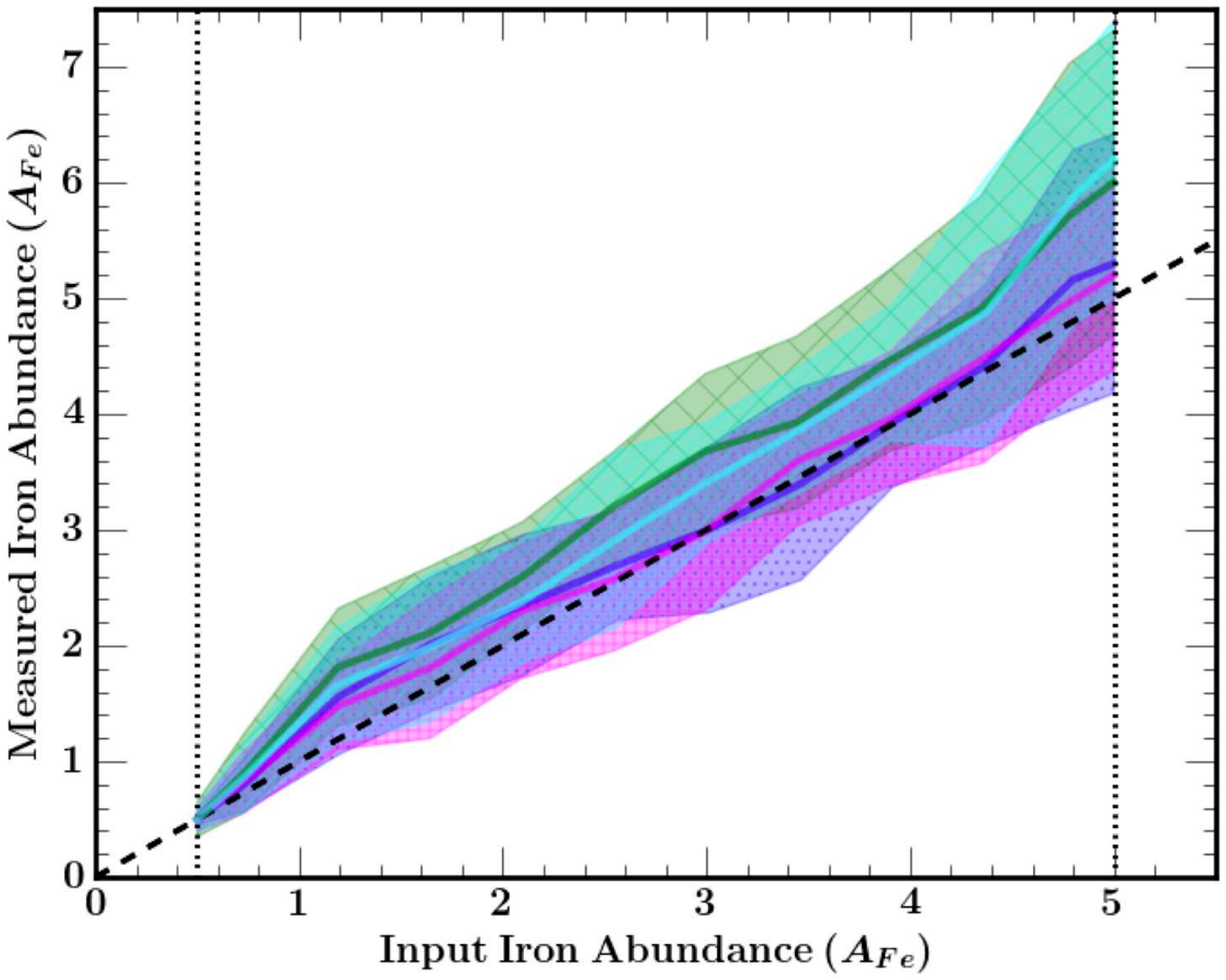}}
	\end{minipage}\\
	\vspace{-2.25in}
	\begin{minipage}{0.5\linewidth}
		\scalebox{0.45}{\includegraphics{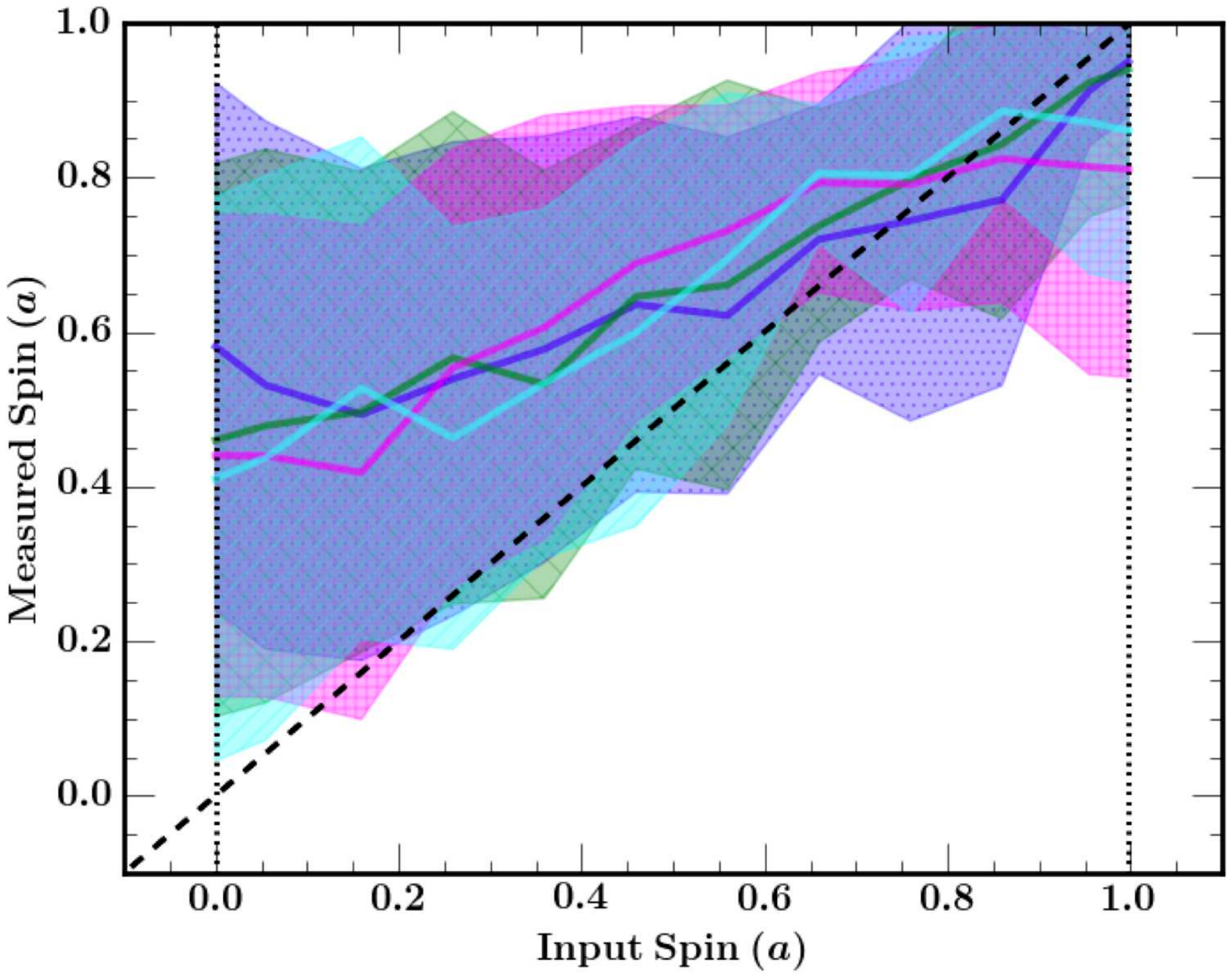}}
	\end{minipage}  
	\begin{minipage}{0.01\linewidth}
		\scalebox{0.45}{\includegraphics{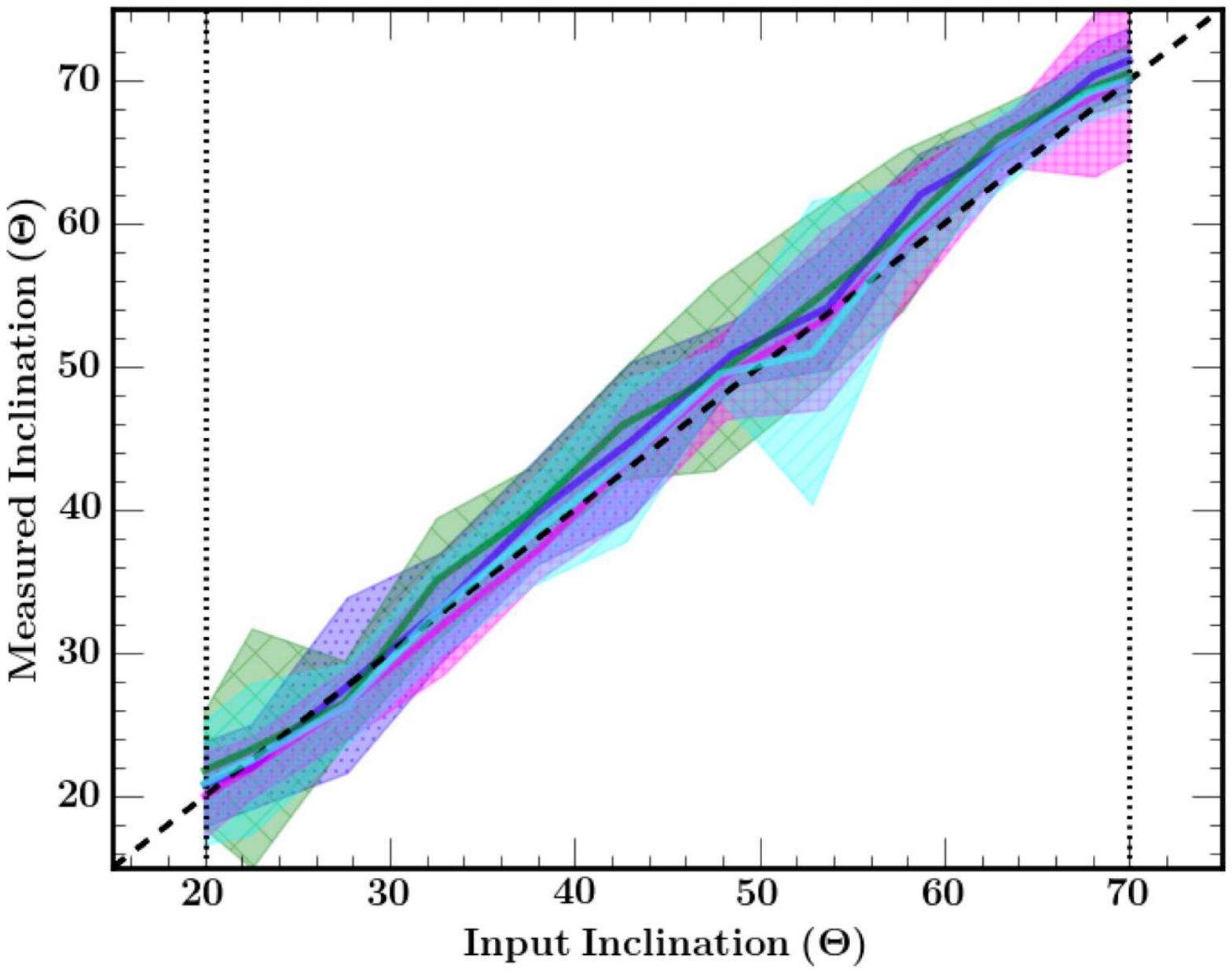}}
	\end{minipage}\\
	\vspace{-2.25in}
	\begin{minipage}{0.5\linewidth}
		\scalebox{0.45}{\includegraphics{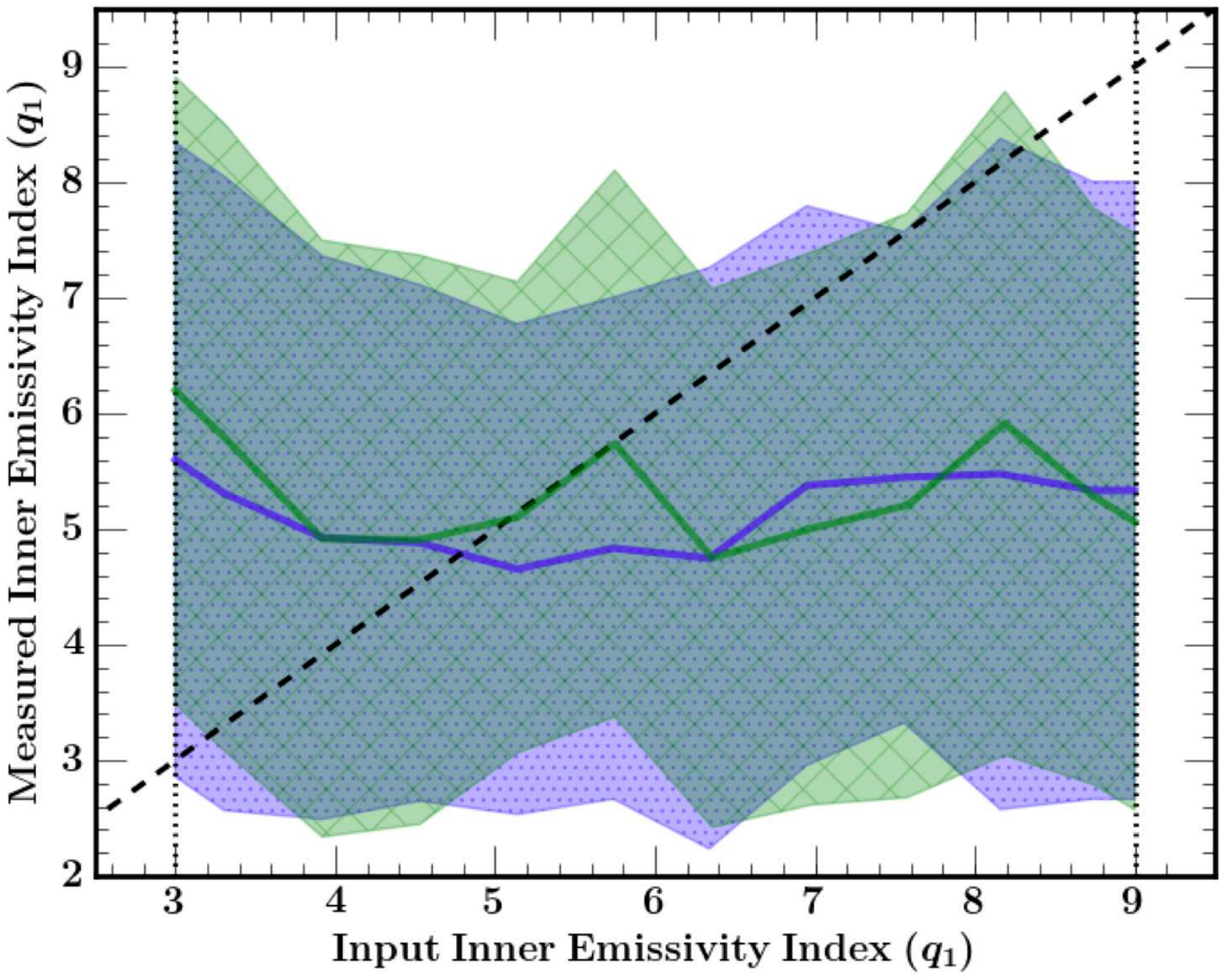}}
	\end{minipage}  
	\begin{minipage}{0.01\linewidth}
		\scalebox{0.45}{\includegraphics{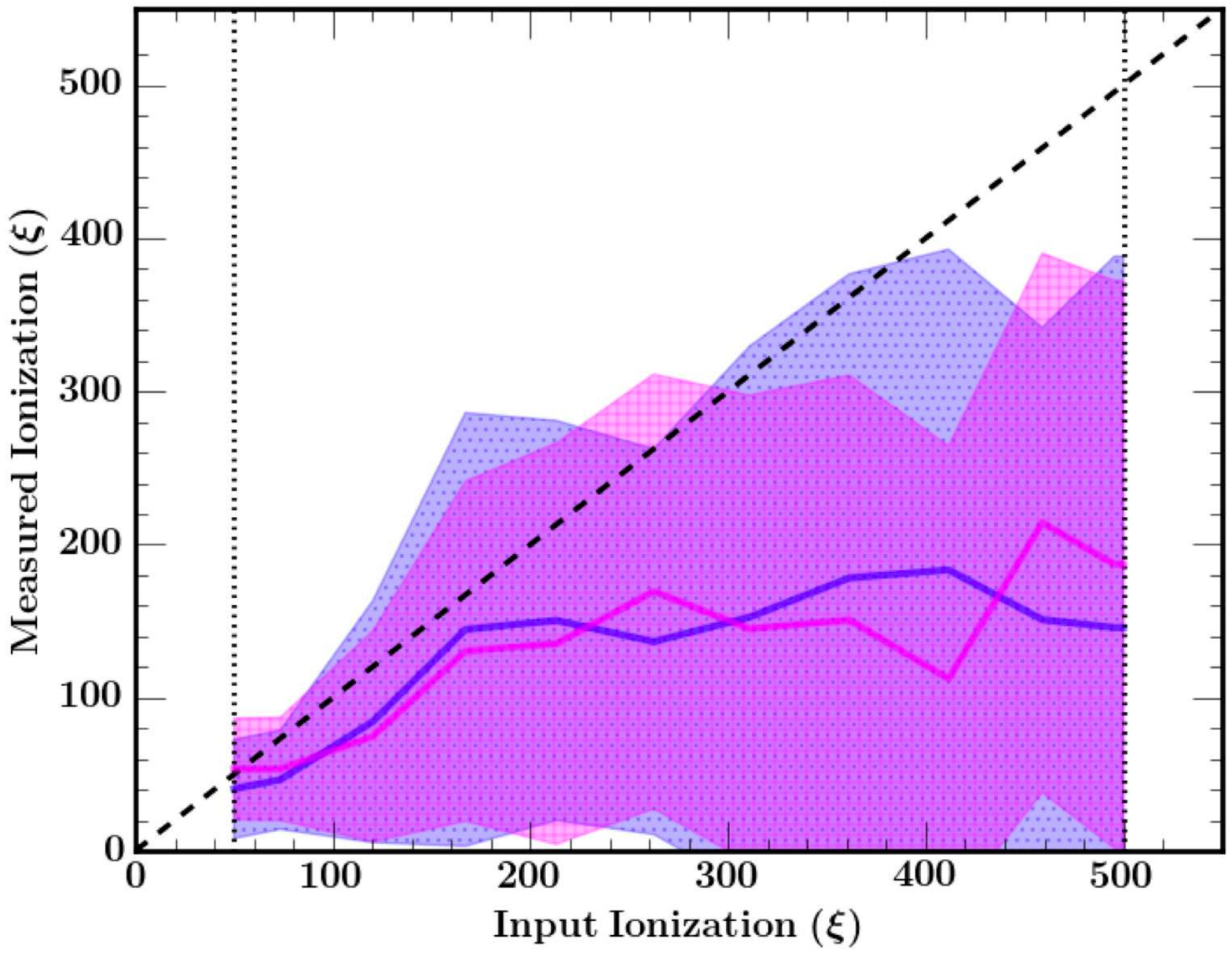}}
	\end{minipage} 
	\vspace{-1in}
    	\caption{Summary of the $R$ = 1 results for all four tests in the 2.5--10\keV\ band. Spectra were binned with respect to input values and plotted with the central solid lines showing the data. The 1$\sigma$ errors for the Test A (small blue dots), Test B (large green hexes), Test C (small pink crosses), and Test D (small cyan hexes) data are illustrated as opaque colored bands. Tests where certain parameters remained fixed are not plotted for that respective parameter (e.g. Test C, D for $q_{1}$ and Test B, D for $\xi$).}
   	 \label{R1XMM_bands}
\end{figure*}
%%%%%%%%%%%%%%%%%%%%%%%%%%%%%%%%%%

%%%%%%%%%% Figure 3: R1 XMM + NuStar binned plots %%%%%%%%%
\begin{figure*}
	\centering
	\vspace{-1in}
	\advance\leftskip-3.5in   
	\begin{minipage}{0.5\linewidth}
		\scalebox{0.45}{\includegraphics{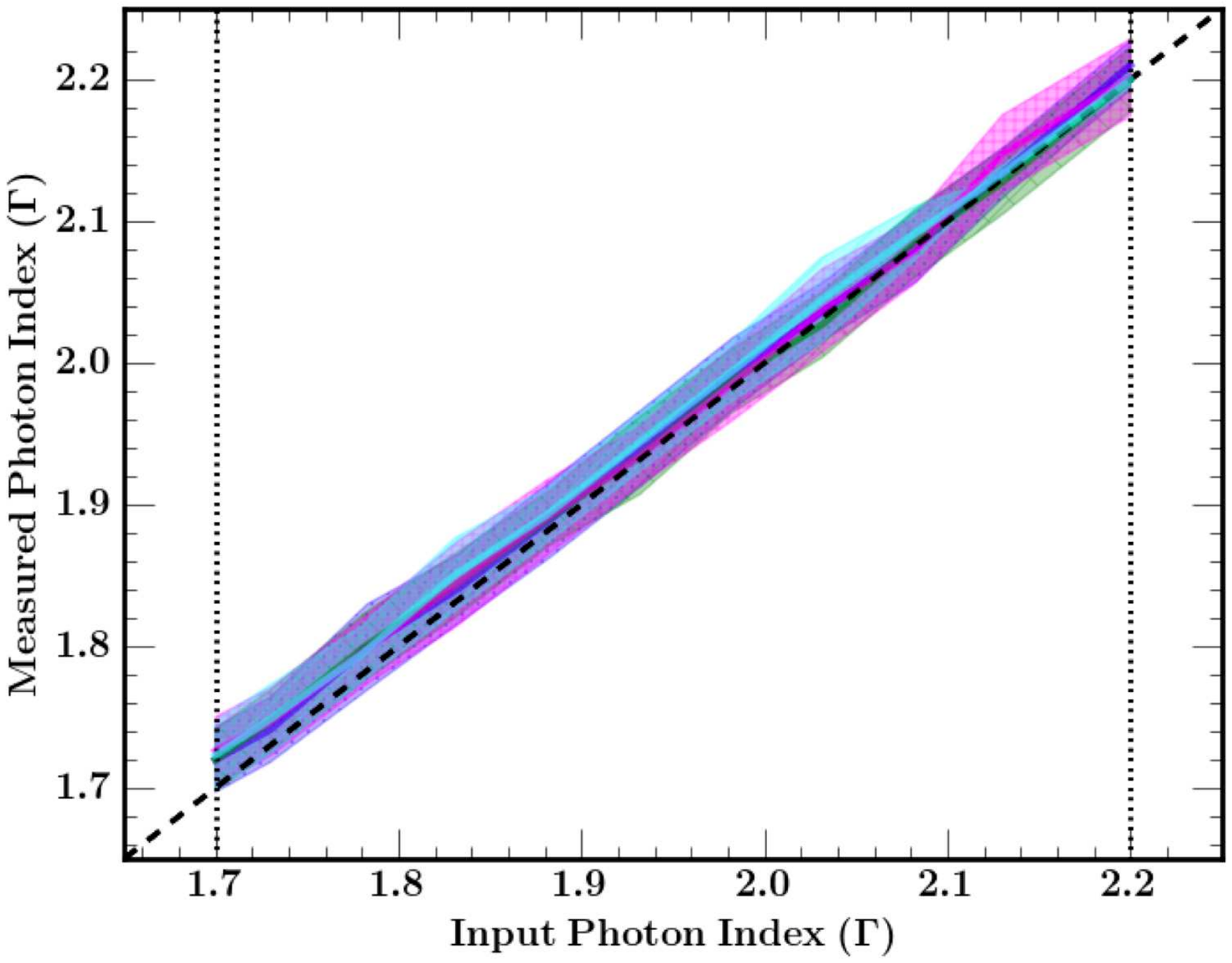}}
	\end{minipage} 
	\begin{minipage}{0.01\linewidth}
		\scalebox{0.45}{\includegraphics{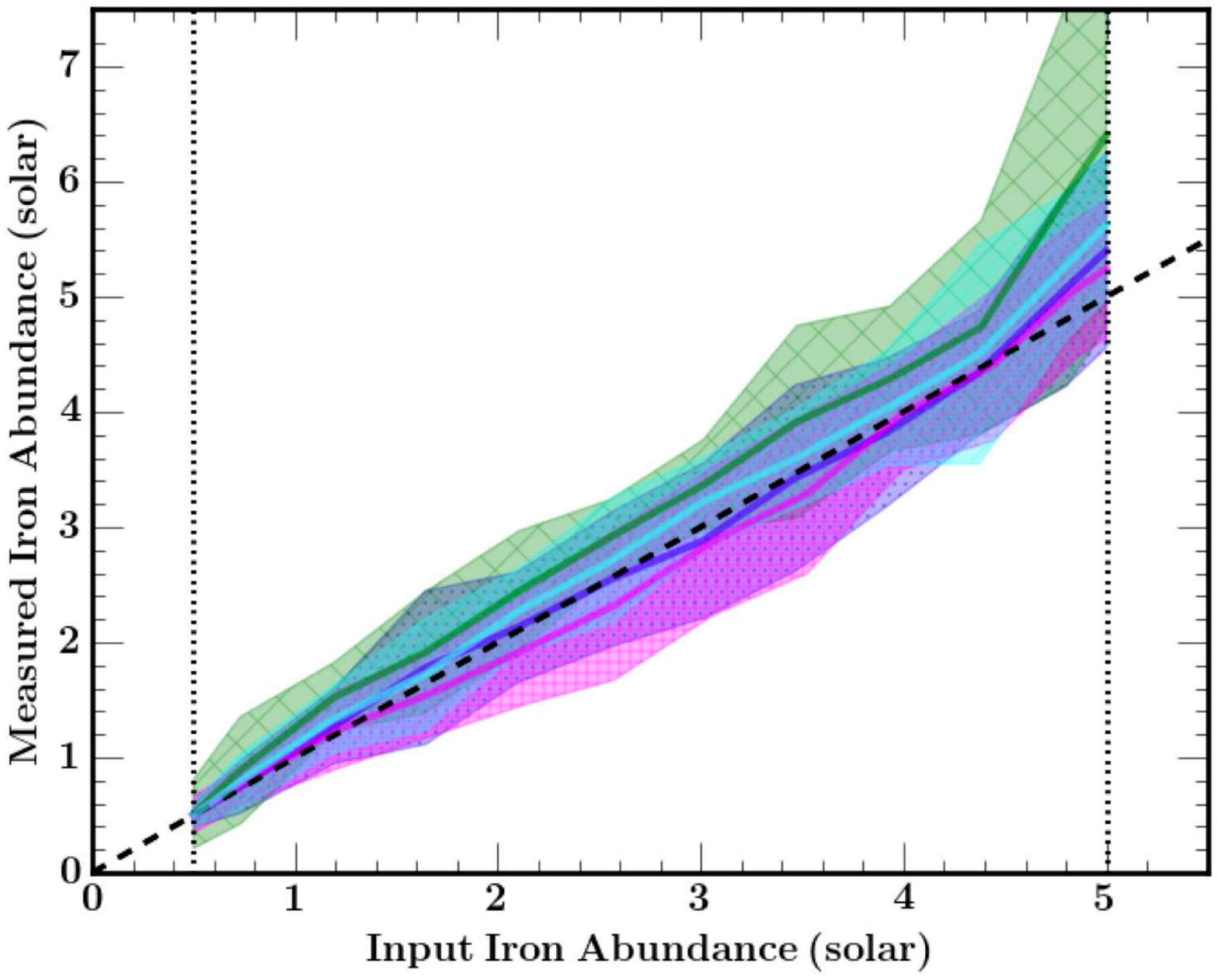}}
	\end{minipage}\\ 
	\vspace{-2.25in}
	\begin{minipage}{0.5\linewidth}
		\scalebox{0.45}{\includegraphics{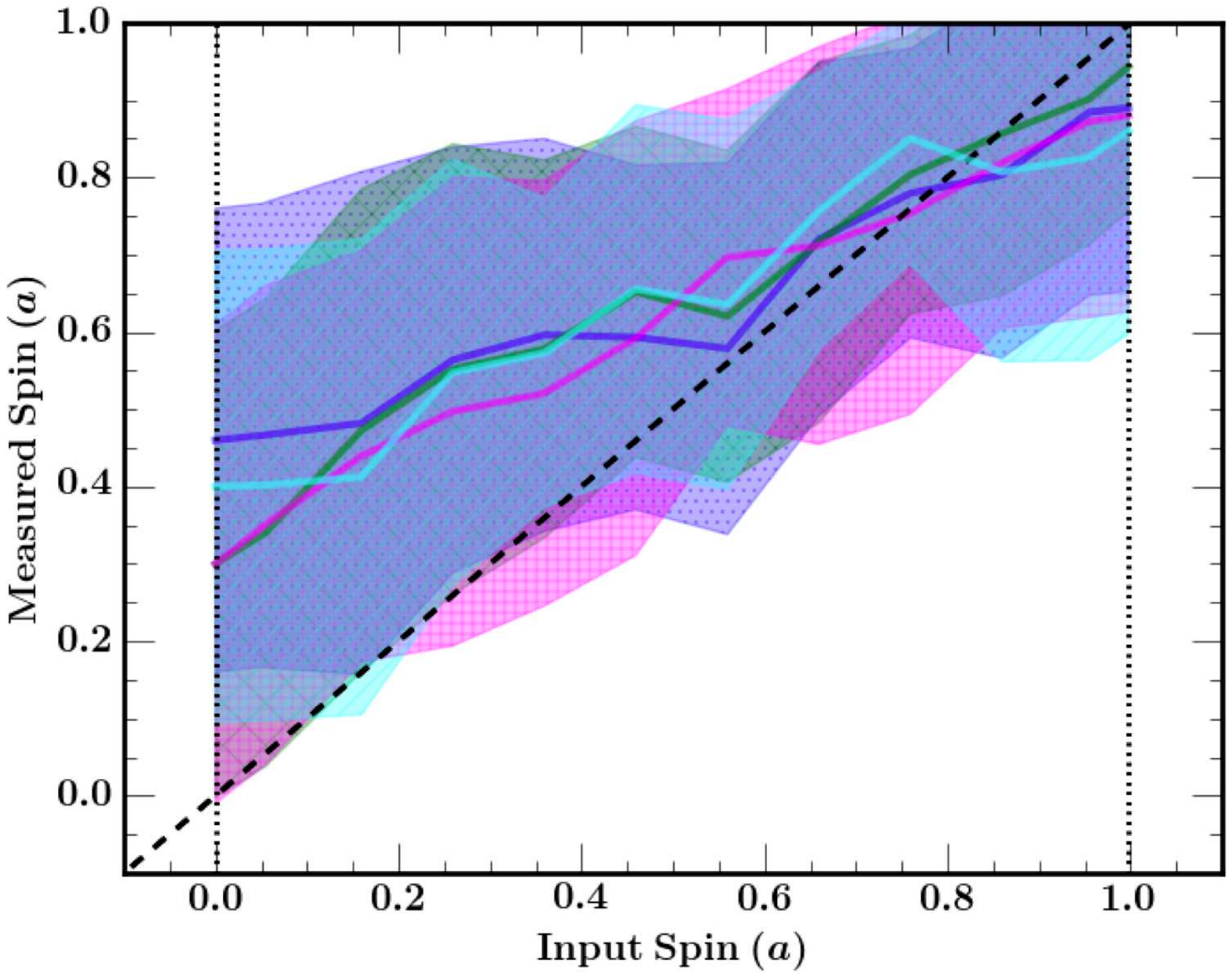}}
	\end{minipage}  
	\begin{minipage}{0.01\linewidth}
		\scalebox{0.45}{\includegraphics{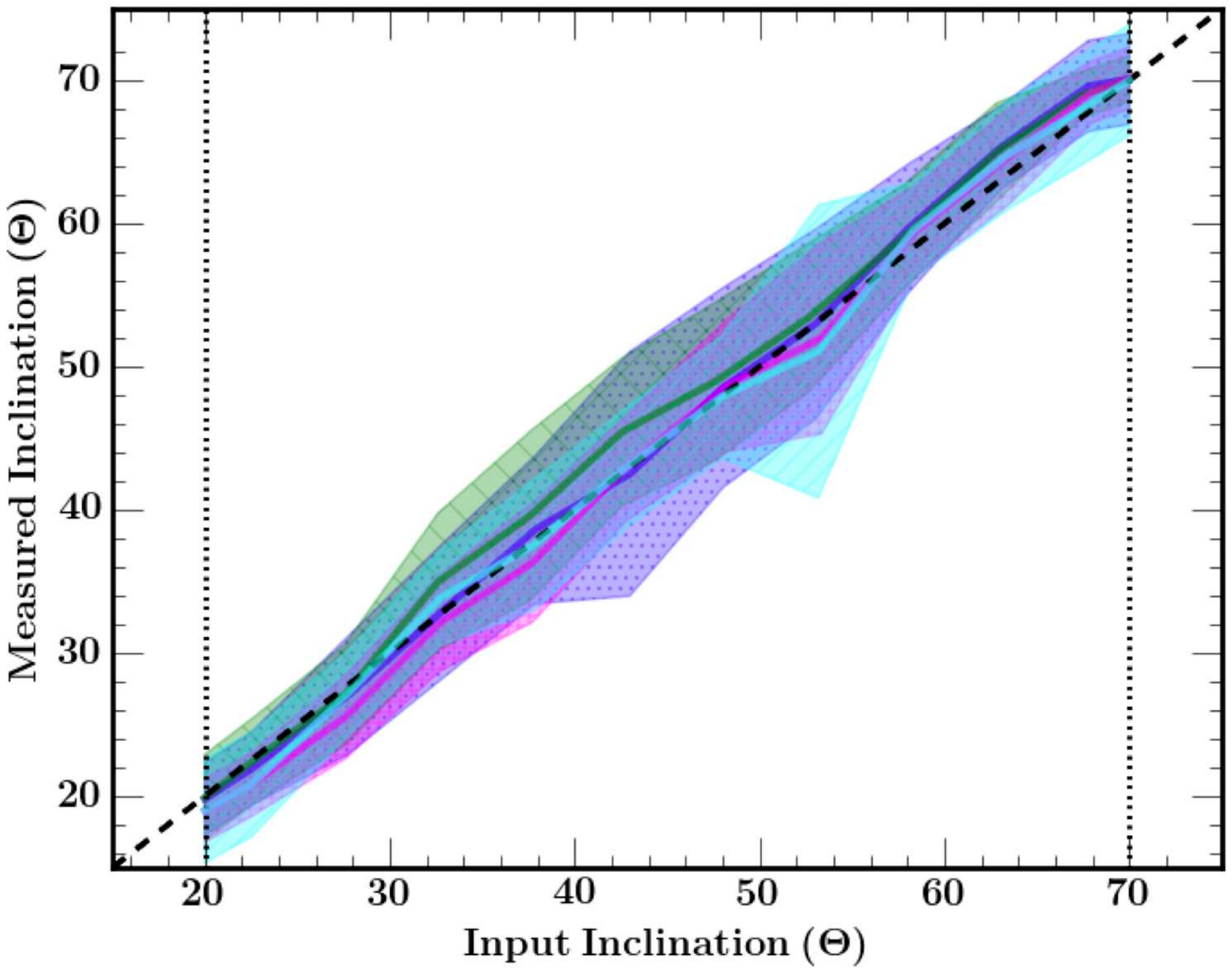}}
	\end{minipage}\\
	\vspace{-2.25in}
	\begin{minipage}{0.5\linewidth}
		\scalebox{0.45}{\includegraphics{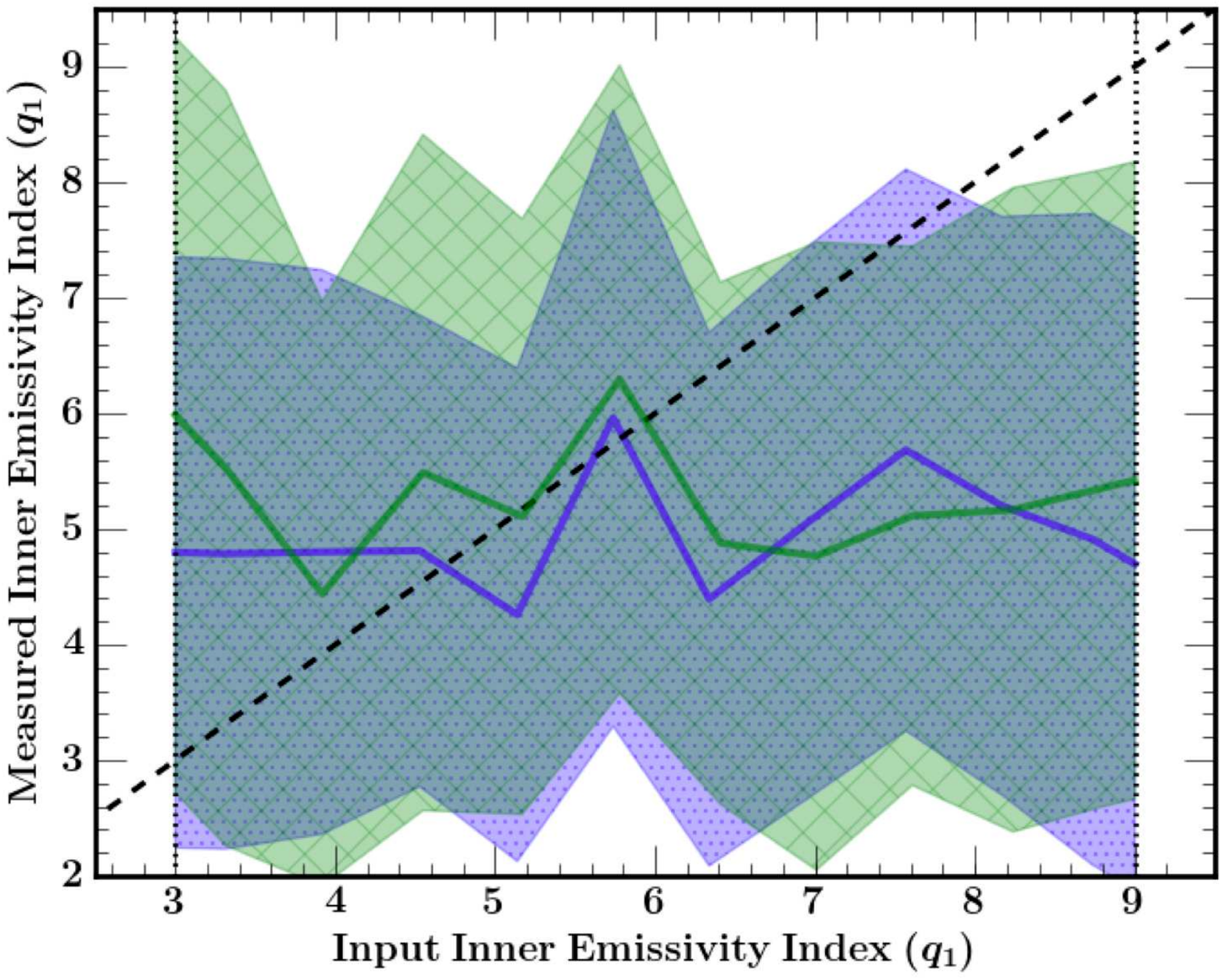}}
	\end{minipage}  
	\begin{minipage}{0.01\linewidth}
		\scalebox{0.45}{\includegraphics{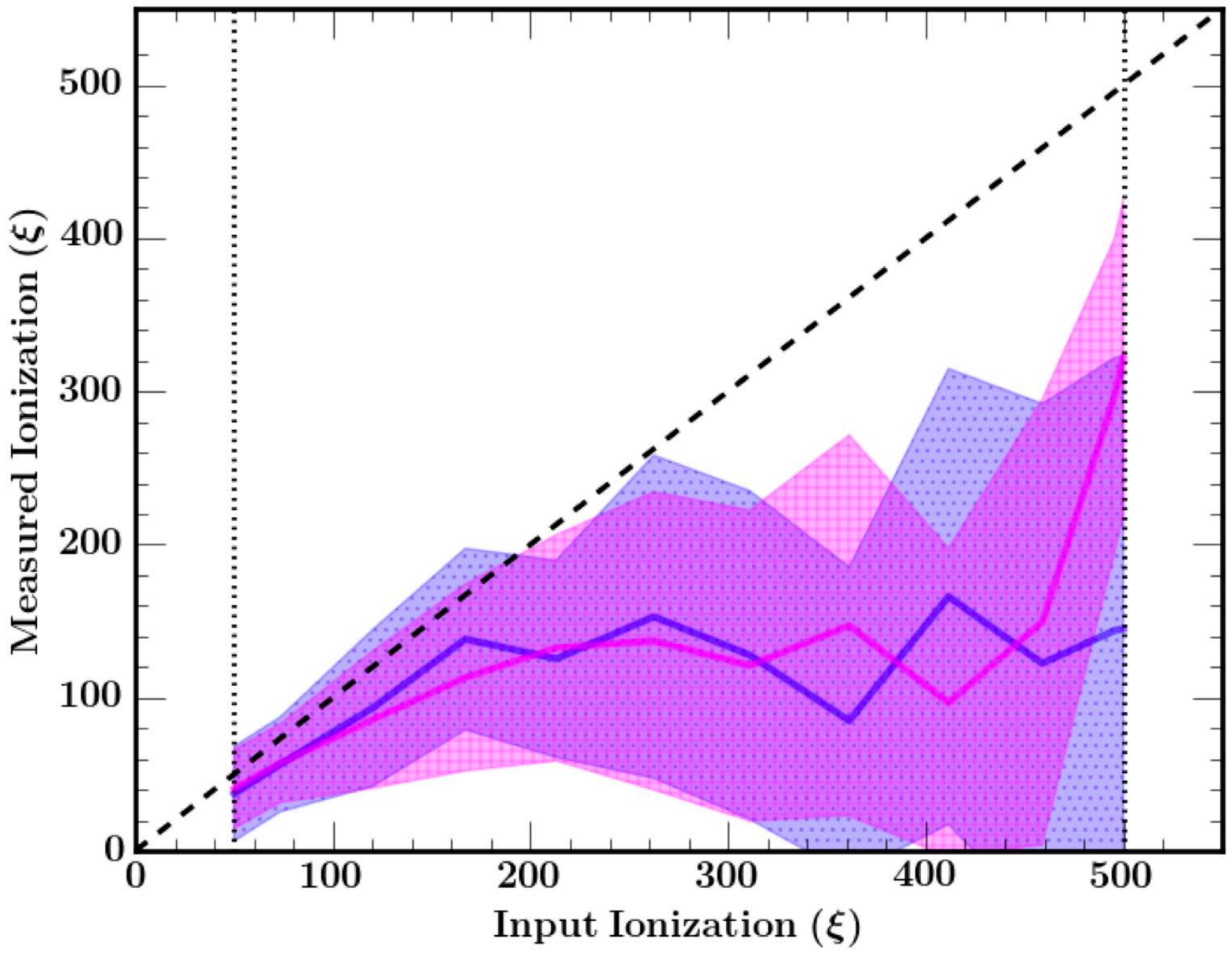}}
	\end{minipage} 
	\vspace{-1in}
    	\caption{Summary of the $R$ = 1 results for the extended 2.5 -- 70\keV\ fit tests. The spectra that were originally produced using \xmm pn response files and fit from 2.5 -- 10\keV\ were simulated once more, with the same input parameter values, using \nustar FMPA and FPMB responses for E$>$10\keV. The spectra were simultaneously refit from 2.5 -- 70\keV. Band colors and patterns are the same as those in Fig. \ref{R1XMM_bands}.}
   	 \label{R1XMM2N_bands}
\end{figure*}
%%%%%%%%%%%%%%%%%%%%%%%%%%%%%%%%%%

%------------------------------------------------------------------------------------

\section{Results from a Reflection Fraction of 1}
\label{R1}
All plot results show a comparison of measured parameter values versus those simulation input values (i.e. ``intrinsic'' spectral values). In order to quantify a models ability to return the input parameters, one can compare the spread in input values for a given measured value. For example, in Fig. \ref{R1XMMcombo} the plot of photon index shows that a measured value of $\Gamma$ = 1.9 indicates a possible input value of between $\Gamma_{in}$ = 1.85 -- 1.92, giving a range of $\sim$0.07 in $\Gamma_{in}$. It is this \emph{input value spread} that we use to visualize how well an AGN parameter can be reproduced.

\subsection{$R$ = 1: 2.5 -- 10 keV spectral fits}
\label{R1_2to10}
The results of the simulated spectral analysis from the 2.5 -- 10\keV\ fitting of Test A are shown in Fig.\thinspace\ref{R1XMMcombo}. The measured vs. input parameter values are plotted in comparison to the 1:1 dashed line denoting a ``correct'' measurement. Dotted vertical lines indicate the ranges in which the random input parameters were generated. Each black point is a simulated AGN spectrum that was autonomously modelled as described in Section \ref{sim_details} and has a chi-squared fit statistic \redchisq $<$ 1.1. To better visualize the results, the data were then binned by input value and overplotted (Fig.\thinspace\ref{R1XMM_bands}). The center points of these binned data are illustrated by the solid lines with the corresponding shaded bands showing the 1$\sigma$ error. 

Trends are immediately apparent through simple visual inspection of the colored bands in each panel. For example, the photon index is measured with the most precision (spread in $\Gamma_{in}$ ranging $\sim$0.06), but is consistently overestimated by a few percent of the correct value. Inclination is measured reasonably well ($\theta_{in}$ ranging $\sim$10\deg) and iron abundance is also well constrained ($A_{Fe,in}$ ranging $\sim$1.6 solar), if slightly overestimated. Spin parameter initially appears to be more difficult to measure for AGN with $R$ = 1 using a limited bandpass. Measured values only begin to converge for $a$ $>$ 0.9. The fit model seems insensitive to both $\xi$ and $q_{1}$.  A detailed discussion of spin will be saved for Section \ref{discussion}.

Lastly, there are few differences between the model fit tests. Test A, B, C, and D all appear to have similar results and none provides a clear advantage over the others with regards to accurately measuring spectral parameters.

\subsection{$R$ = 1: 2.5 -- 70 keV spectral fits}
\label{R1_2to70}
The binned results of the extended energy band fitting are shown in Fig.\thinspace\ref{R1XMM2N_bands}. We can visually confirm that photon index is now both accurately measured and tightly constrained. Iron abundance results are also consistent with those of the 2.5 -- 10\keV\ band, although when the ionization parameter, $\xi$, is kept fixed (Test B, large green hexes) the measured values are systematically over-estimated. Inclination angle remains well constrained, especially for input values $\theta$ $<$ 30\deg\ and $>$ 60\deg\ where $\theta_{in}$ ranges $\sim$8\deg, for an improvement over the narrower spectral analysis (Section \ref{R1_2to10}) by $\sim$2\deg. Inner emissivity and ionization continue to be unconstrained parameters. Once again, there is no significant difference between the model fit tests.

%------------------------------------------------------------------------------------

\section{Results for a Reflection Fraction of 5} 
\label{R5}
\subsection{$R$ = 5: 2.5 -- 10 keV spectral fits}
\label{R5_2to10}
\begin{figure*}
	\centering
	%\minipage{1.0\pagewidth}
		\vspace{-1in}
		\advance\leftskip-3.5in   
		\begin{minipage}{0.5\linewidth}
			\scalebox{0.45}{\includegraphics{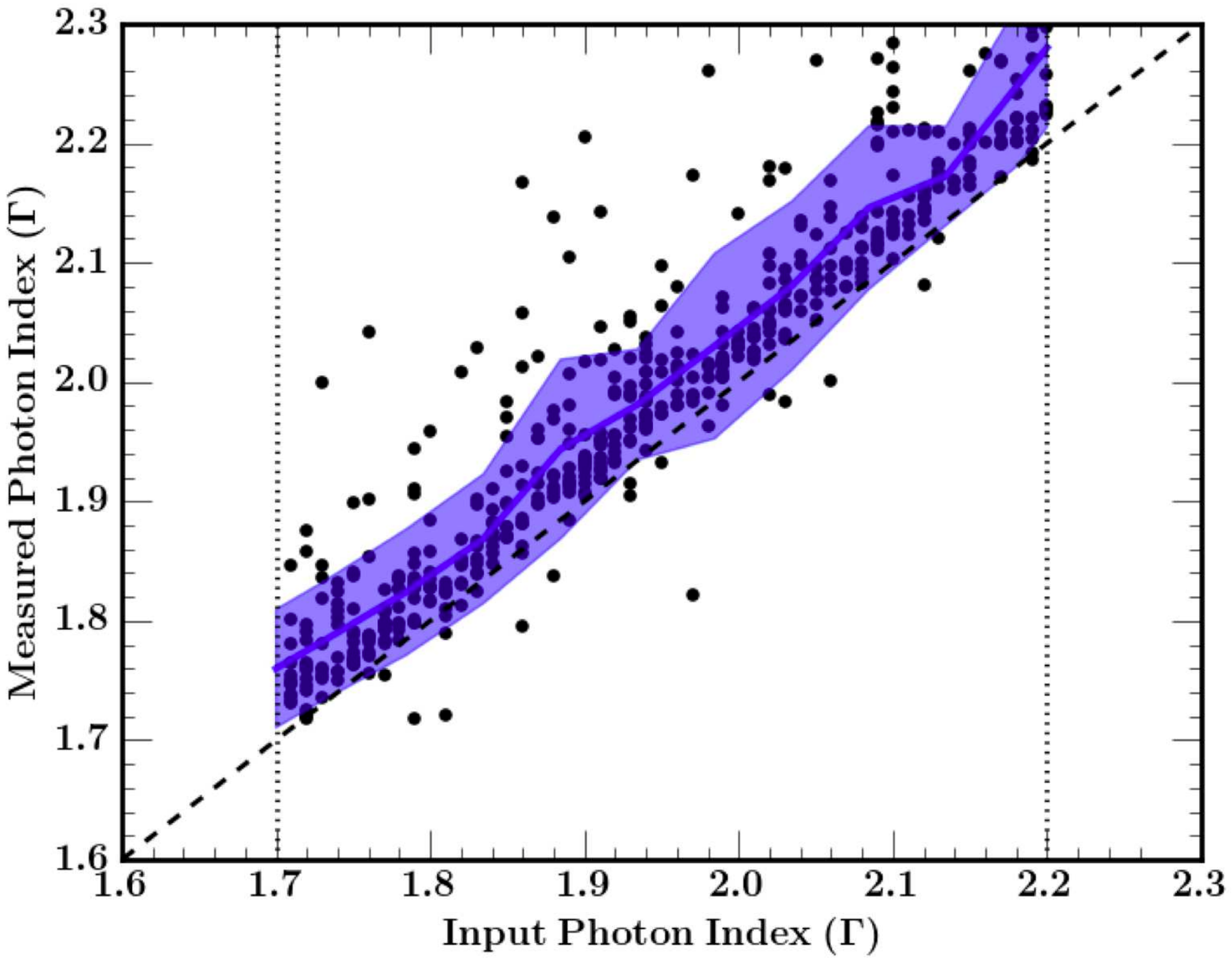}}
		\end{minipage} 
		\begin{minipage}{0.01\linewidth}
			\scalebox{0.45}{\includegraphics{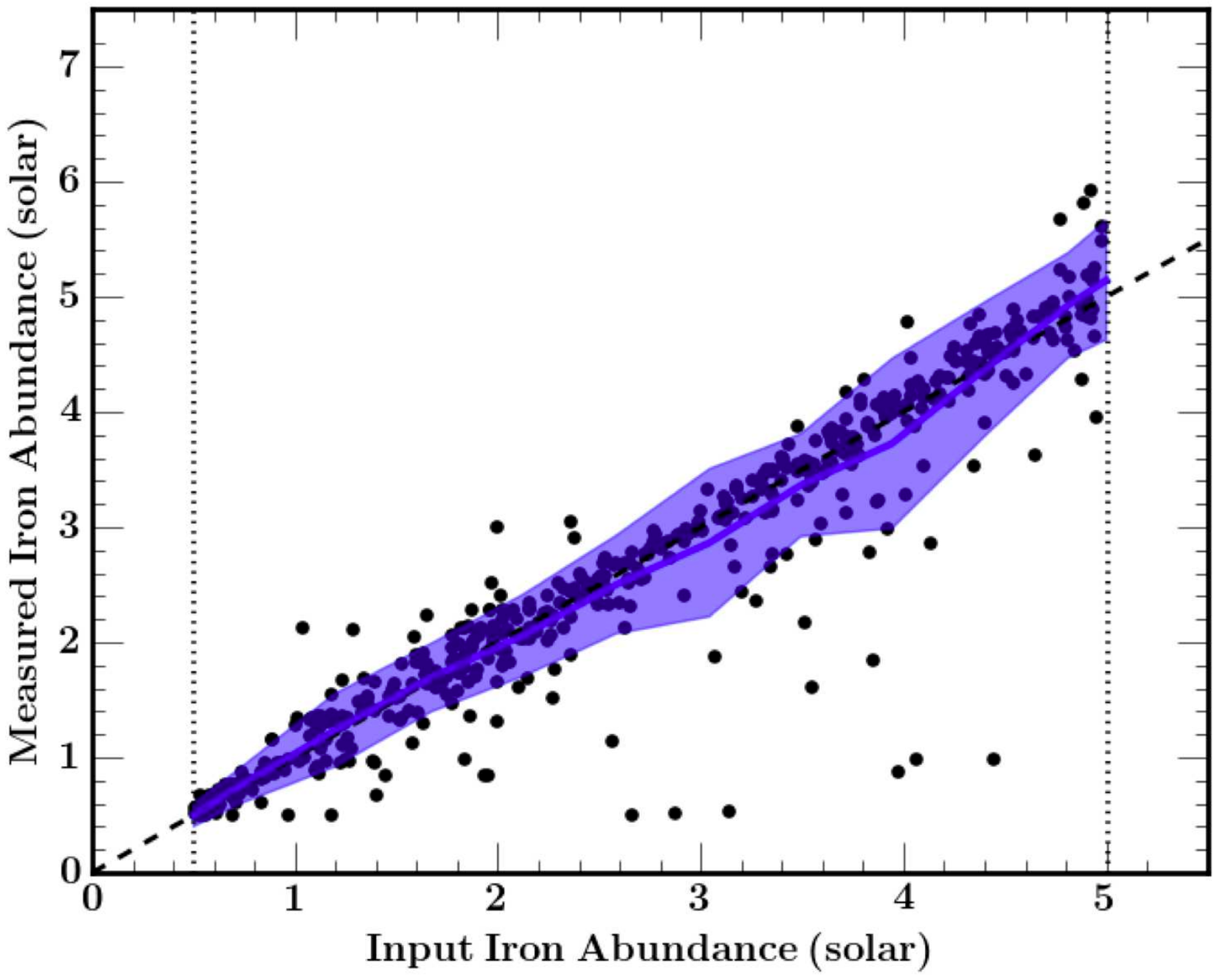}}
		\end{minipage}\\ 
		\vspace{-2.25in}
		\begin{minipage}{0.5\linewidth}
			\scalebox{0.45}{\includegraphics{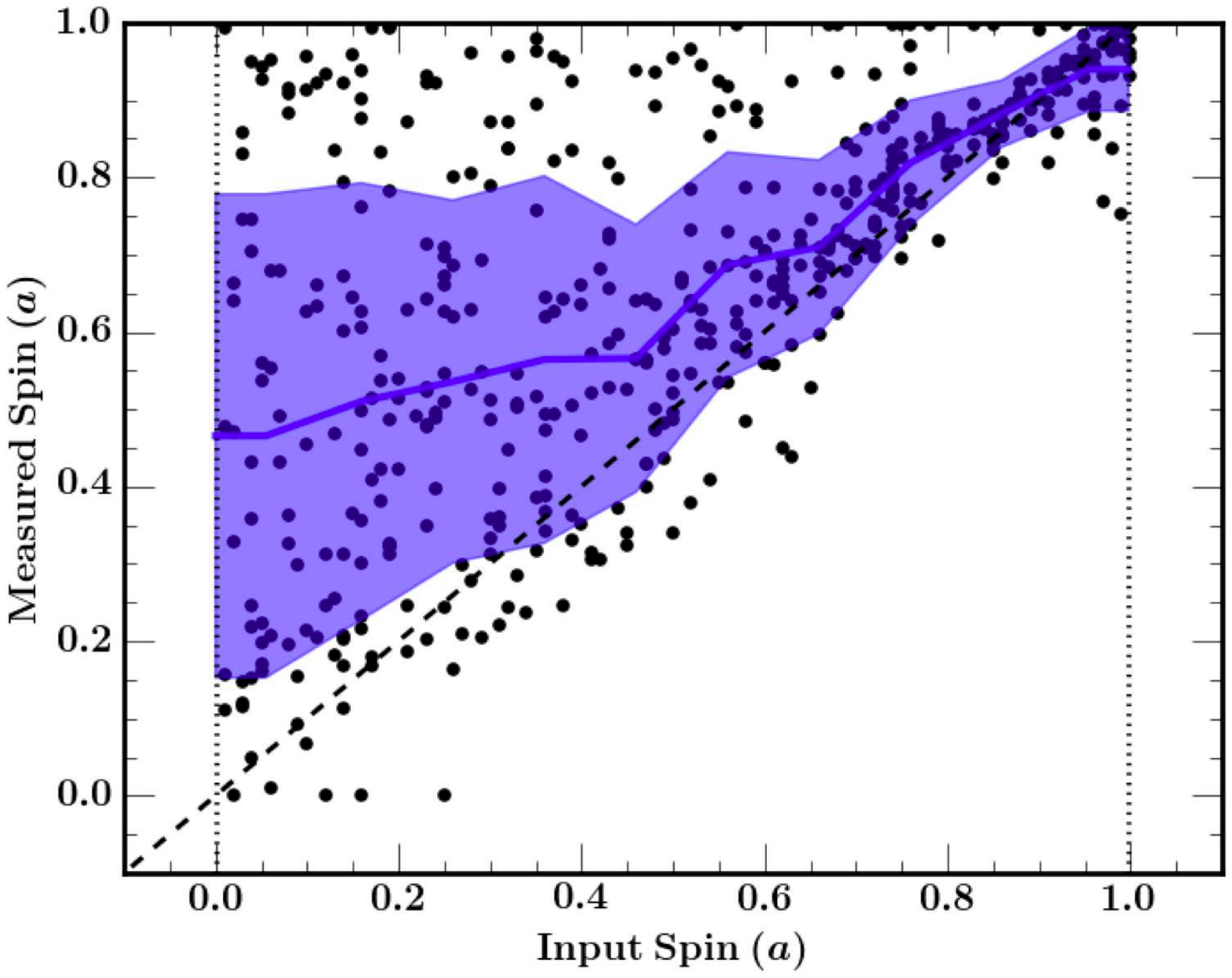}}
		\end{minipage}  
		\begin{minipage}{0.01\linewidth}
			\scalebox{0.45}{\includegraphics{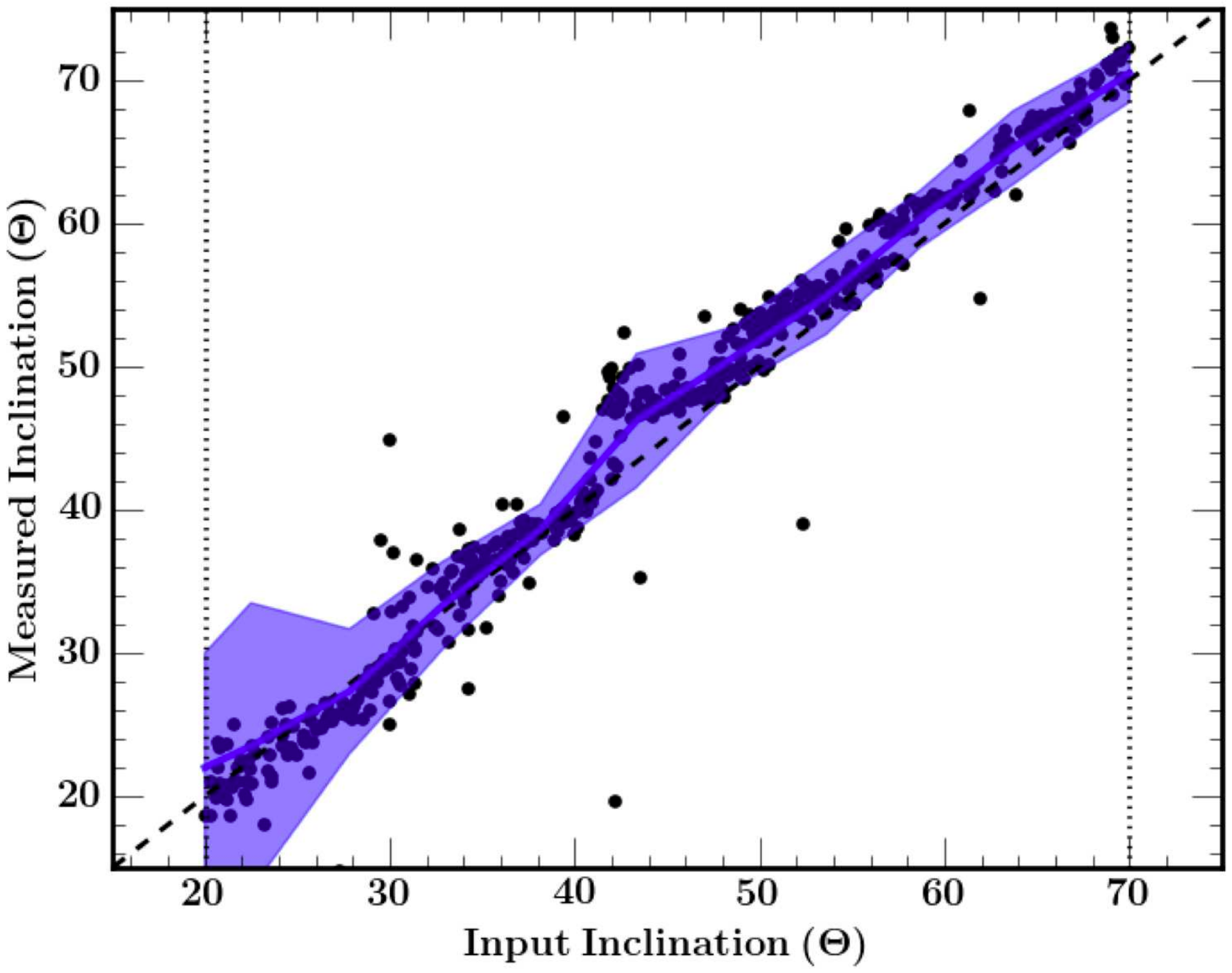}}
		\end{minipage}\\
		\vspace{-2.25in}
		\begin{minipage}{0.5\linewidth}
			\scalebox{0.45}{\includegraphics{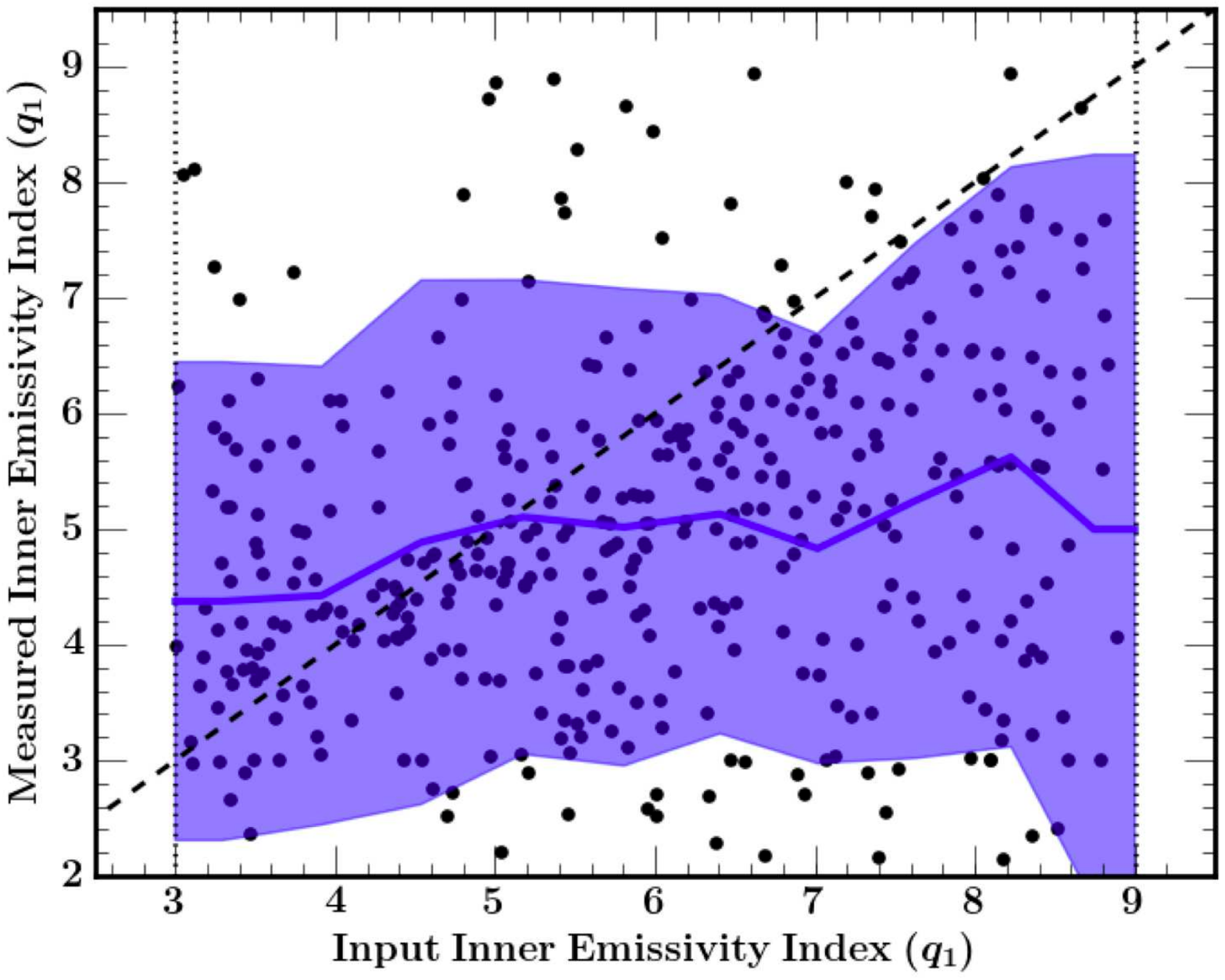}}
		\end{minipage}  
		\begin{minipage}{0.01\linewidth}
			\scalebox{0.45}{\includegraphics{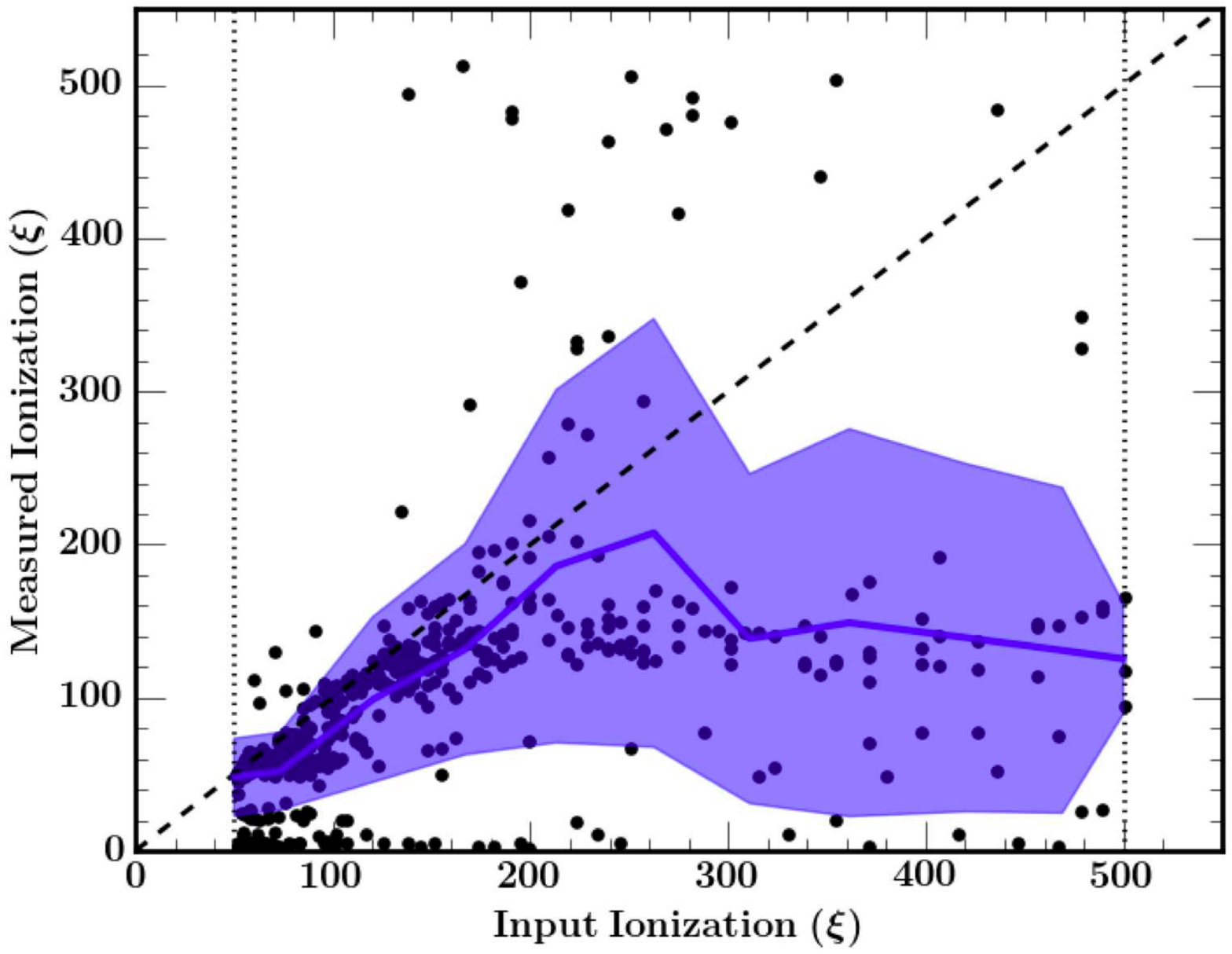}}
		\end{minipage} 
		\vspace{-1in}
	%\end{minipage}
    	\caption{The $R$ = 5 simulated spectral fitting from 2.5--10\keV. Only the results from Test A, where all 6 parameters were free to vary, are shown for simplicity. Plot details are the same as for those in Fig.\thinspace\ref{R1XMMcombo}. As expected, an increase in reflection fraction decreases measured precision in primary continuum parameter $\Gamma$, but increases measured precision and accuracy in reflection parameters $A_{Fe}$, $a$, and $\theta$.}
   	 \label{R5TAcombo}
\end{figure*}

The results of the simulated spectral analysis from the 2.5 -- 10\keV\ fitting of Test A are shown in Fig.\thinspace\ref{R5TAcombo}; graph details are the same as those in Fig.\thinspace\ref{R1XMMcombo}. As expected, our ability to measure photon index decreases --- shown by the parameter being more over-predicated as compared to the 2.5 -- 10\keV\ results for $R$ = 1 and the standard deviations also increasing. Measured iron abundance and inclination angle become more precise for most measured values, with $A_{Fe,in}$ range decreasing to $\sim$0.8 solar and $\theta_{in}$ range decreasing to $\sim$5\deg. However, $A_{Fe}$ is slightly underestimated at intermediate values and $\theta$ remains slightly overestimated throughout. Spin is now significantly better constrained as $a$ increases ($a_{in}$ ranging $\sim$0.1 for a measured value of $a$ = 0.95), so it appears an increase in reflection fraction does indeed influence our ability to constrain it. Lastly, ionization also seems to be significantly better constrained than in the $R$ = 1 scenario, for $\xi_{in}$ $<$ 250\ergcmps, above which the parameter is once again unconstrained. This could arise from line profile diminishing with increased ionization. Inner emissivity index ($q_{in}$) remains unconstrained.

As with the $R$ = 1 analysis, there are few differences between the individual fit tests and we show Test A results for the 2.5 -- 10\keV\ band in Fig.\thinspace\ref{R5TAcombo} as an accurate representation of all four.

\subsection{$R$ = 5: 2.5 -- 70 keV spectral fits}
\label{R5_2to70}
\begin{figure*}
	\centering
	\vspace{-1in}
	\advance\leftskip-3.5in   
	\begin{minipage}{0.5\linewidth}
		\scalebox{0.45}{\includegraphics{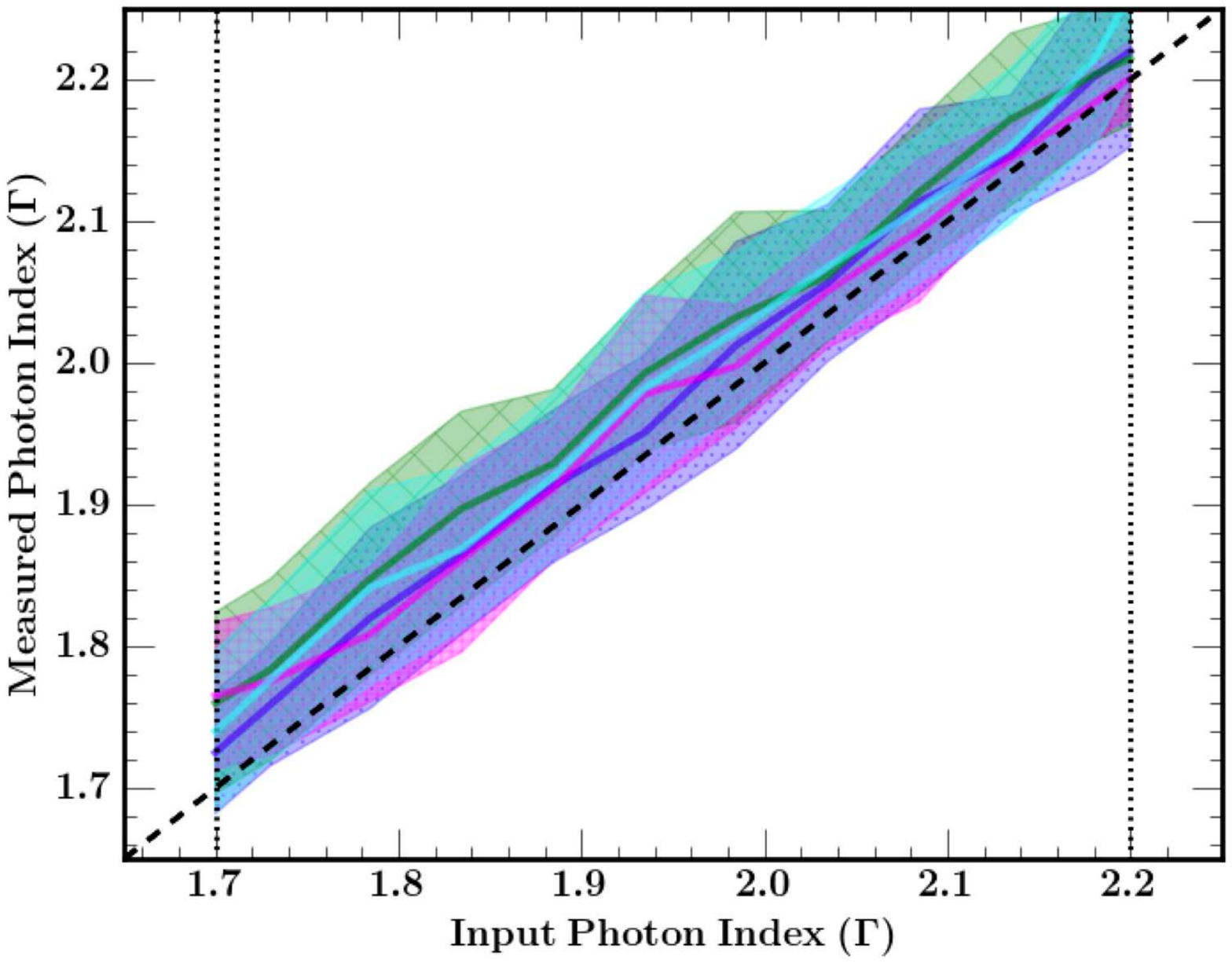}}
	\end{minipage} 
	\begin{minipage}{0.01\linewidth}
		\scalebox{0.45}{\includegraphics{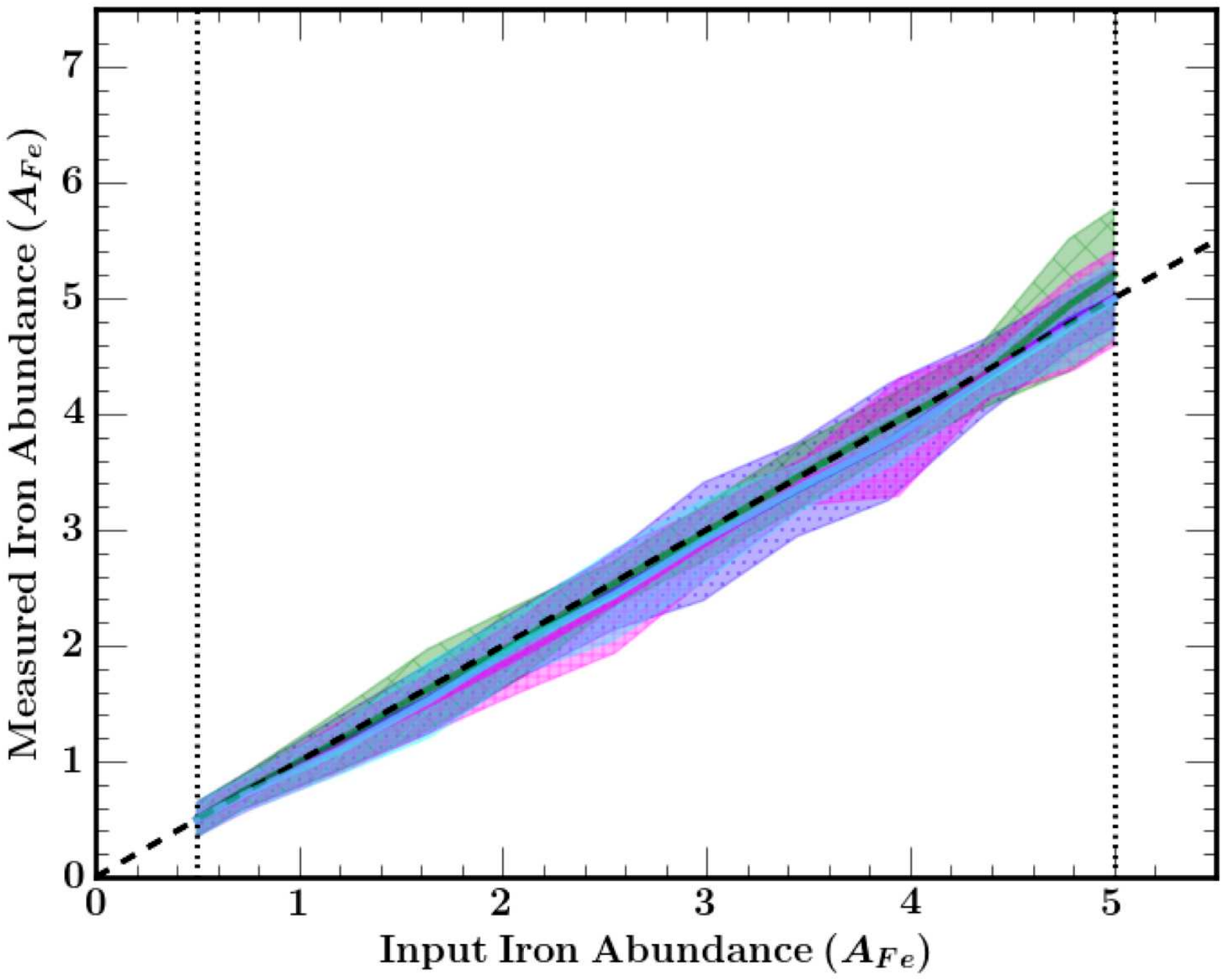}}
	\end{minipage} \\
	\vspace{-2.25in}
	\begin{minipage}{0.5\linewidth}
		\scalebox{0.45}{\includegraphics{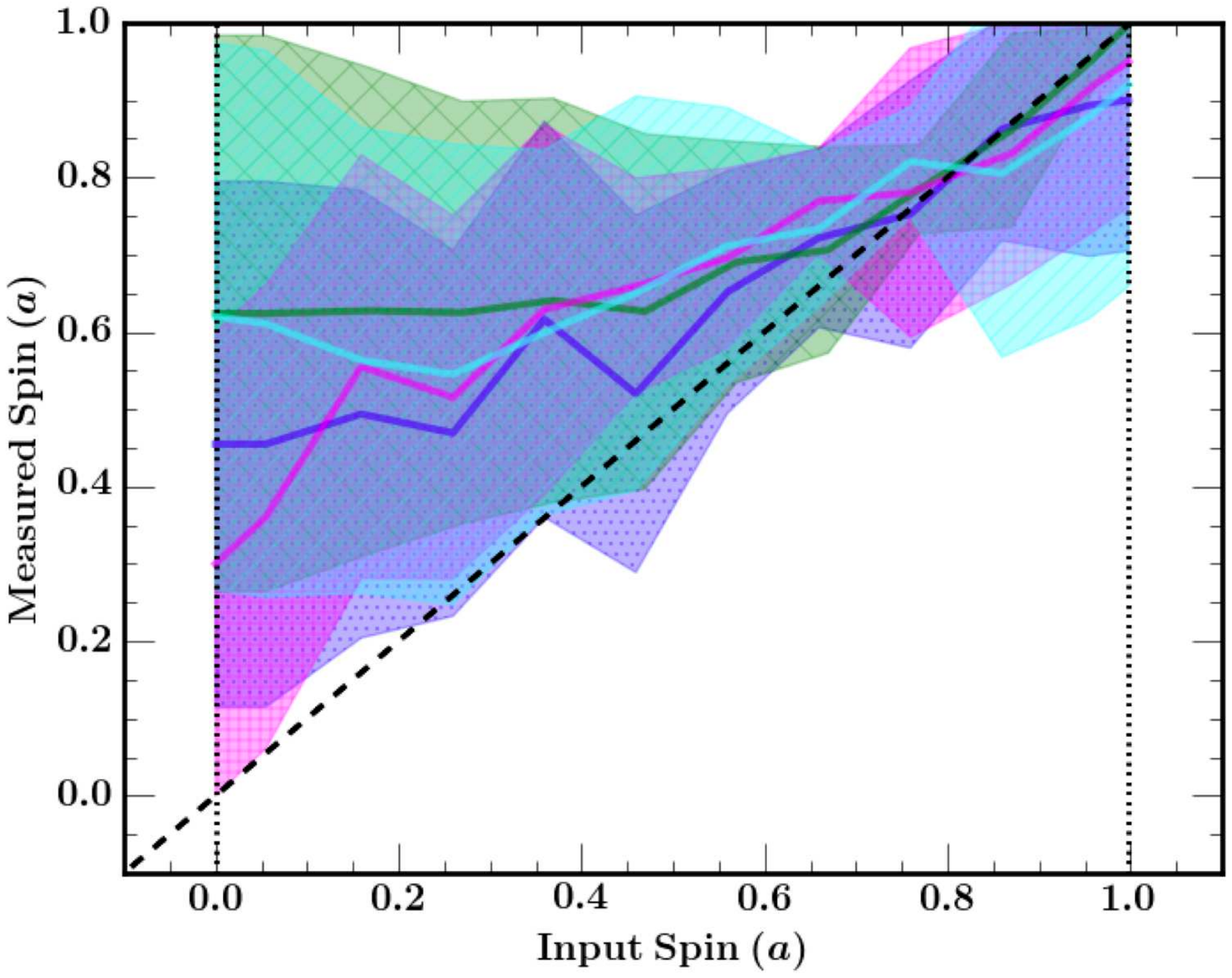}}
	\end{minipage}  
	\begin{minipage}{0.01\linewidth}
		\scalebox{0.45}{\includegraphics{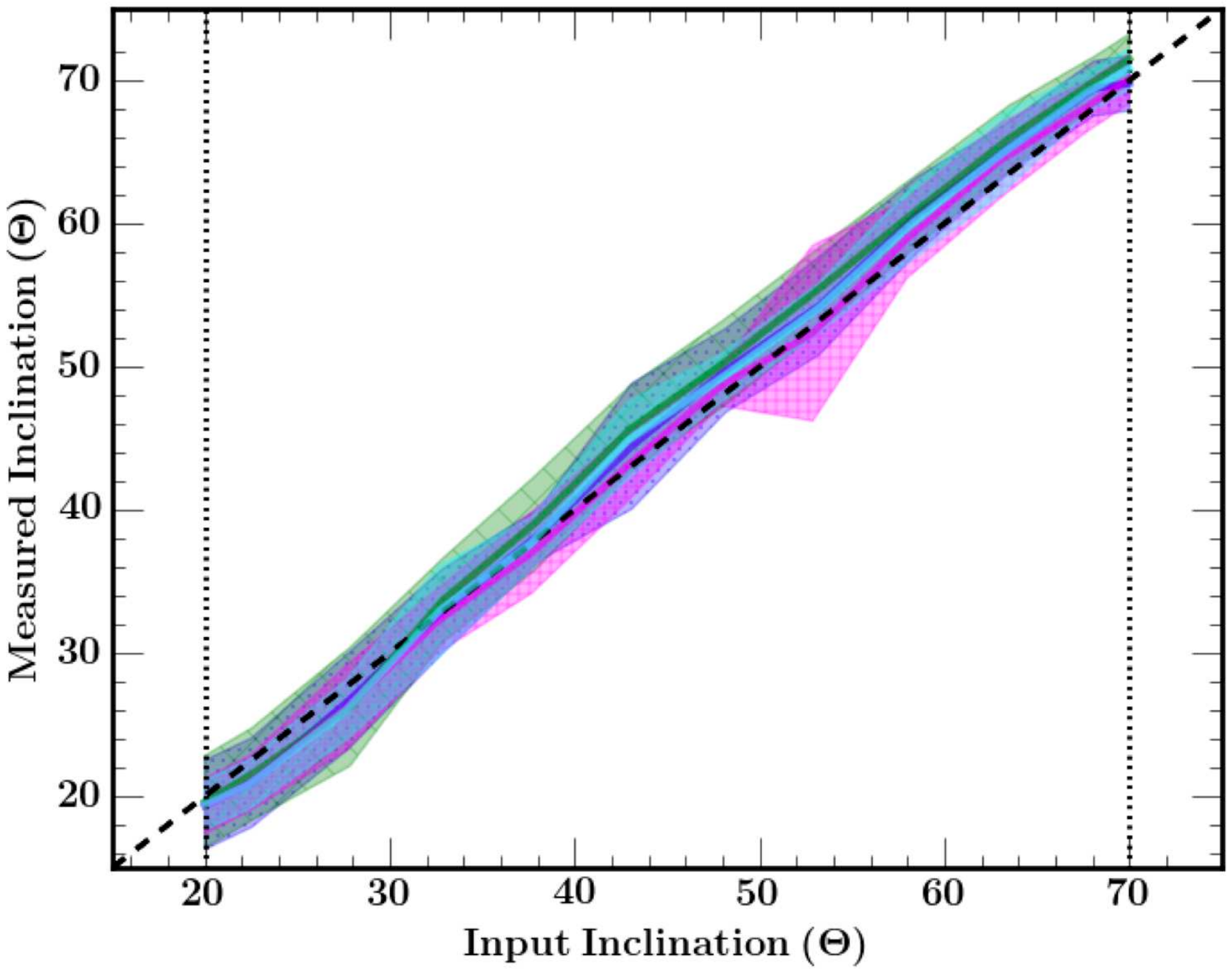}}
	\end{minipage}\\
	\vspace{-2.25in}
	\begin{minipage}{0.5\linewidth}
		\scalebox{0.45}{\includegraphics{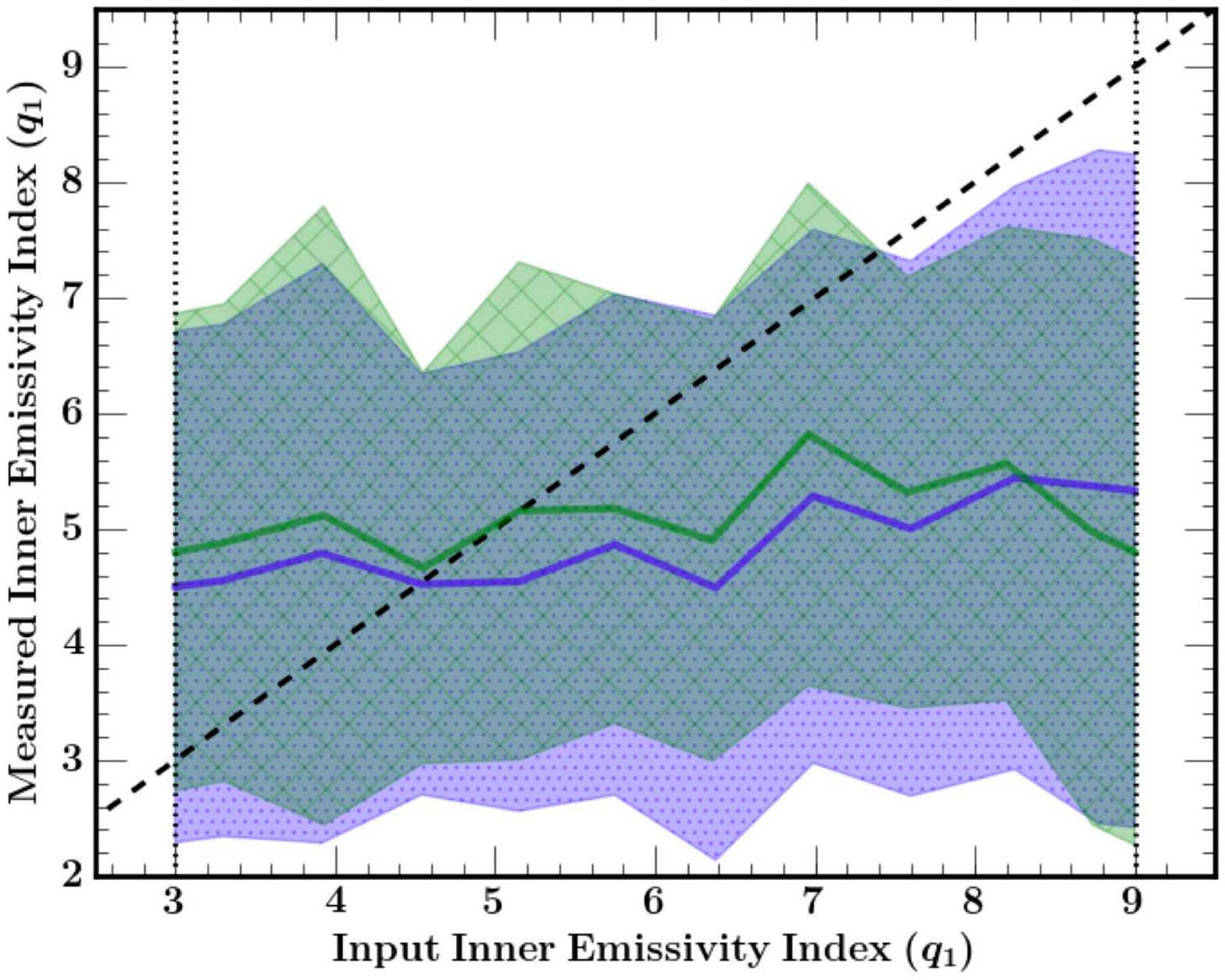}}
	\end{minipage}  
	\begin{minipage}{0.01\linewidth}
		\scalebox{0.45}{\includegraphics{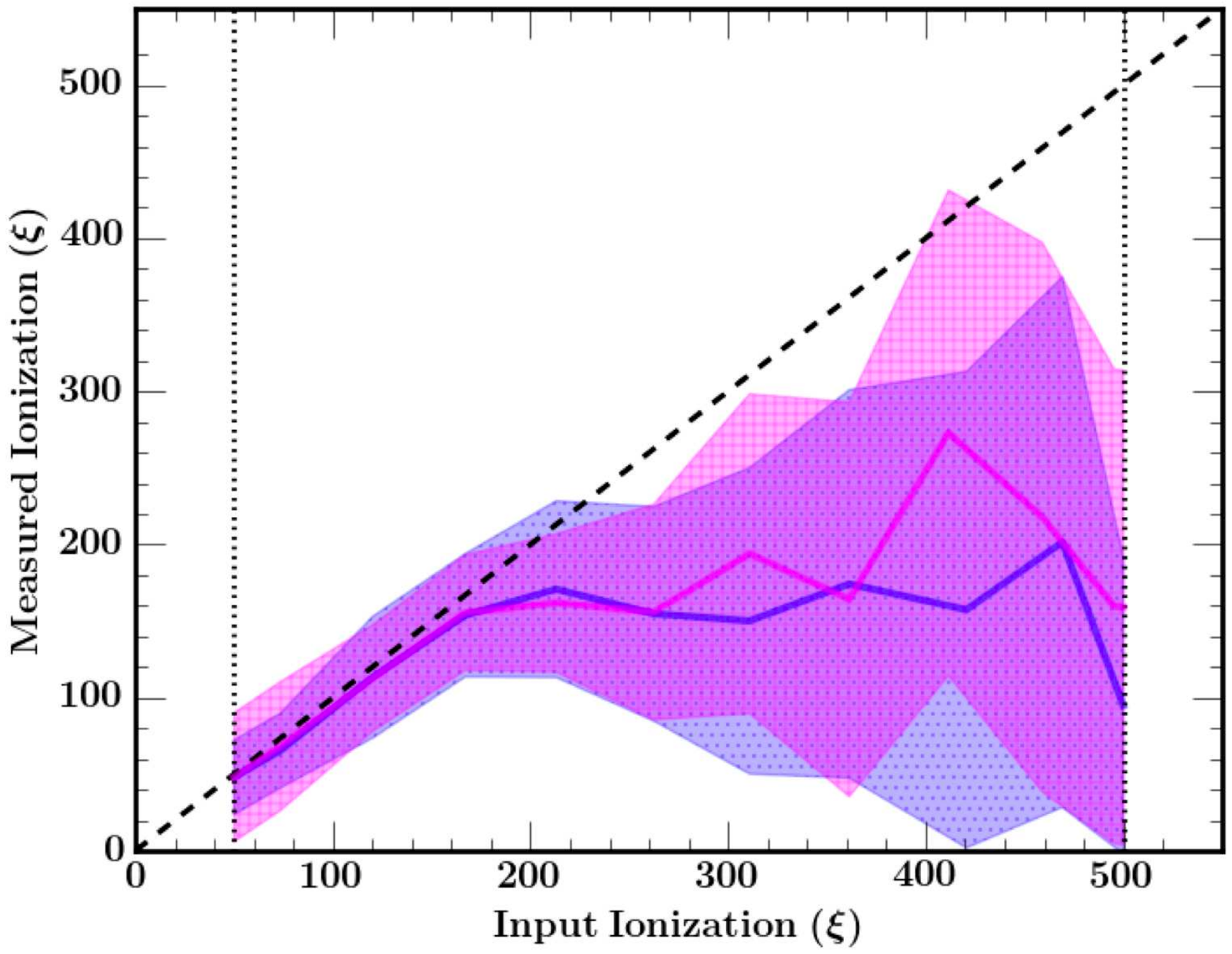}}
	\end{minipage} 
	\vspace{-1in}
    	\caption{Summary of the $R$ = 5 results for the extended 2.5--70\keV\ fit tests. Figure details are the same for those in Fig.\thinspace\ref{R1XMM_bands}. When spectral fits are extended up to 70\keV, parameters $A_{Fe}$, $\theta$, and $\xi$ are better constrained. However, it appears we do not improve our ability to measure $a$. Measurements of $a$ continue to be an improvement over those for the $R$ = 1 scenario in the same bandpass.}
   	 \label{R52N_bands}
\end{figure*}

The binned results of the extended energy band fitting are shown in Fig.\thinspace\ref{R52N_bands}; plot details are the same as those in Fig.\thinspace\ref{R1XMM2N_bands}. When the 2.5 -- 10\keV\ $R$ = 5 spectra are refit up to 70\keV, we increase our ability to constrain most reflection parameters. The input range of $A_{Fe,in}$ decreases to $\sim$0.6 solar and we especially improve our ability to constrain the lower measurement limits for a given value of $A_{Fe,in}$. Our range in $\theta_{in}$ remains around 5\deg\ and, for ionization values below $\sim$200\ergcmps, precision in $\xi$ increases with the increased bandwidth. Above $\sim$200\ergcmps\ there is no significant distinction between the fit bands, as is to be expected. Unlike the $R$ = 1 scenario, however, we do not improve our ability to constrain black hole spin by increasing the fit bandpass. For this case of a higher reflection fraction, possible input spin values for a given measurement increase to a range of $a_{in}$ = 0.25 for a measured value of $a$ = 0.95.
 
As with all previous results, there is little difference between the model fit tests for this extended bandpass, however all four fit tests are shown in Fig.\thinspace\ref{R52N_bands} for completeness.

%%%%%%%%%% Figure: XMM only R5TA Combo Scatter + Bin Plots %%%%%%%%%
%%%%%%%%%%%%%%%%%%%%%%%%%%%%%%%%%%

%%%%%%%%%% Figure: R5 XMM + NuStar binned plots %%%%%%%%%
%%%%%%%%%%%%%%%%%%%%%%%%%%%%%%%%%%

\section{Retrograde Spin Investigation} 
\label{retrograde}
\begin{figure*}
	\centering
	\advance\leftskip-0.5in
	\vspace{-1in}
	\advance\leftskip-2.5in   
	\begin{minipage}{0.5\linewidth}
		\scalebox{0.4}{\includegraphics{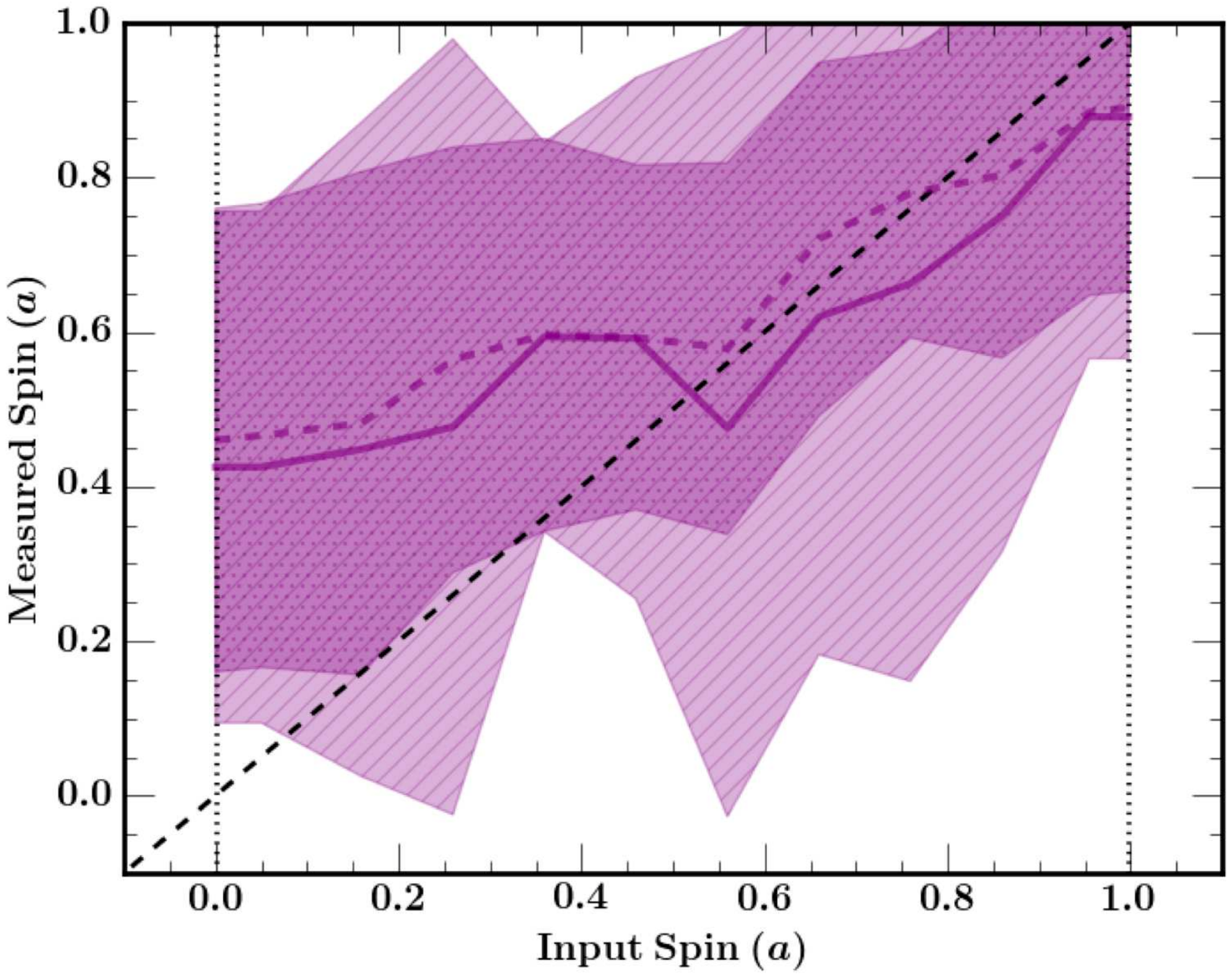}}
	\end{minipage}  
	\begin{minipage}{0.01\linewidth}
		\scalebox{0.4}{\includegraphics{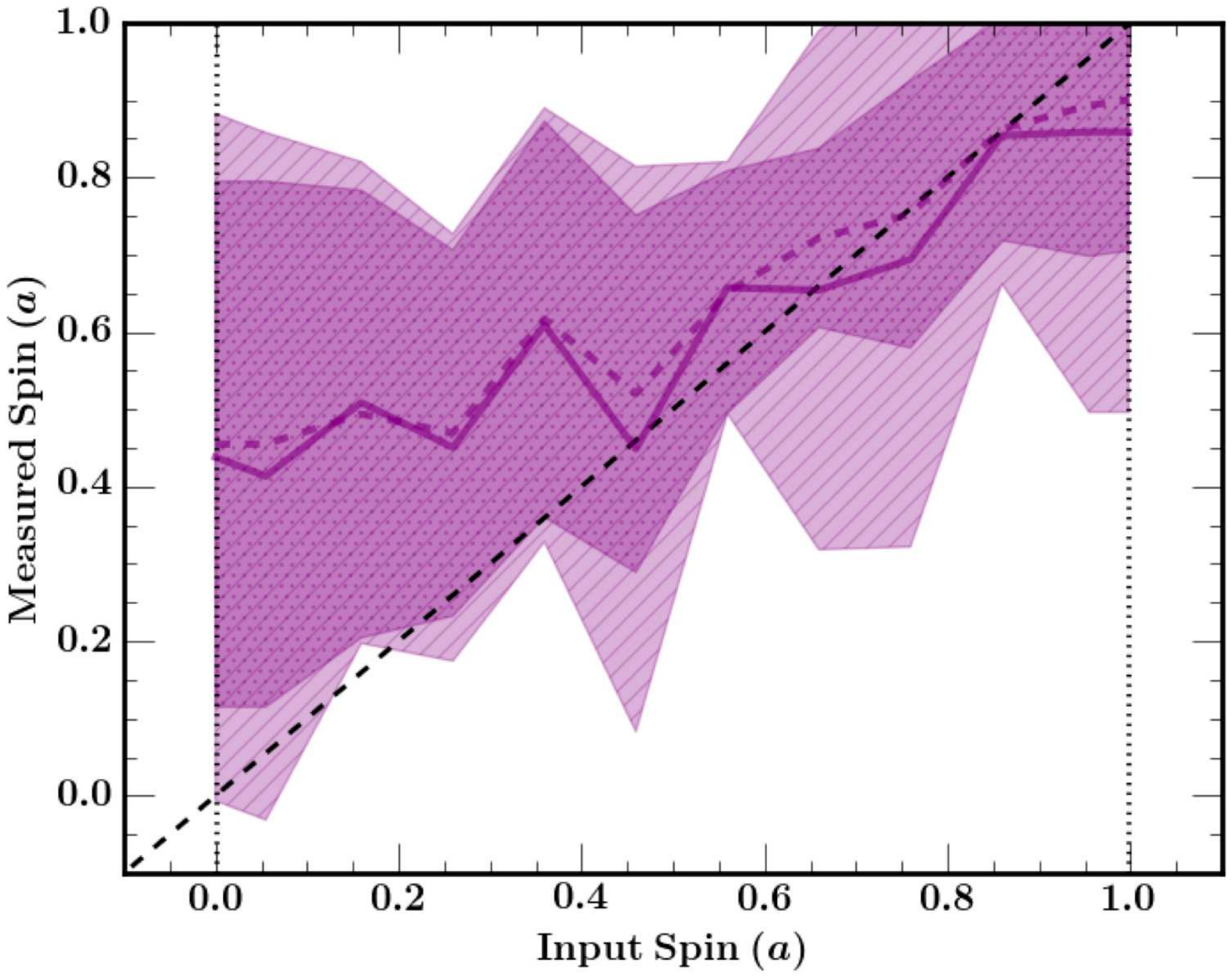}}
	\end{minipage} 
	\vspace{-1in}
    	\caption{Comparison plots of retrograde spin results for the 2--70\keV\ fits for $R$ = 1 (left) and $R$ = 5 (right). The prograde (0$\leq$$a$$\leq$0.998) model fits are illustrated by the dotted line, with a dotted band showing 1$\sigma$ uncertainties. The retrograde (-0.998$\leq$$a$$\leq$0.998) model fits are illustrated by the solid line, with a striped band showing 1$\sigma$ uncertainties. In both cases, allowing the lower limit of model spin to include retrograde fits clearly disrupts our ability to constrain spin at even the highest values of $a$.}
   	 \label{RetSum}
\end{figure*}

Thus far, the possibility of retrograde spin has not been considered. Therefore, the same $R$ = 1 spectra that were used in the 2.5 -- 70\keV\ analysis were refit with the same model and default starting parameters, only now with the spin model boundaries allowing for a retrograde fit (i.e. -0.998 to 0.998). It should be noted that no retrograde fit \emph{should} be found as none of the spectra were simulated with a spin $a$ $<$ 0 (see Section \ref{sim_details}). However, by allowing the model to include retrograde spins in the statistical fitting process, we can investigate any duplicity in measured spin results and their cause (like in the case of 3C~120). Once the lower limit for possible model $a$ values was extended, our ability to constrain even the most extreme spins diminished (Fig. \ref{RetSum}, left). 

Repeating the above procedure for the $R$ = 5 scenario, none of the key parameters were significantly better constrained when the model spin lower limit was relaxed to include a search for a retrograde-spinning black hole (Fig. \ref{RetSum}, right). Photon index remained over-estimated and spin itself was entirely unconstrained. As was the case when $R$ = 1, it appears that including retrograde fits increases the standard deviation of measurements at both min- and maximum spins. Allowing a fit model to process the full possible spin range appears to reduce our ability to measure even the most extreme spin values as tightly.

%%%%%%%%%% Figure: Retrograde Summary %%%%%%%%%
%%%%%%%%%%%%%%%%%%%%%%%%%%%%%%%%%%

Despite the difficulty in constraining spin when allowing a full range of black hole spin values, our ability to measure other parameters remained relatively unchanged. We continue to measure prograde spins, which is to be expected given our sample of exclusively prograde objects, and we do not adversely influence our ability to constrain other key parameters. By including retrograde spins in the modelling we simply reduce our ability to constrain the highest $a$-values.

%------------------------------------------------------------------------------------
 
 \section{Discussion} 
\label{discussion}
\subsection{$\bm{R}$1: band comparison}
\label{disc_R1}
Looking exclusively at the 2 -- 10\keV\ fit results in the $R$ = 1 scenario, photon index and black hole spin tend to be overestimated while observation angle, and iron abundance are well constrained. Photon index is the most consistent parameter and we can be confident that our measurements of $\Gamma$ are accurate to within about 5 percent. Constraints of $\theta$ are accurate overall: a single measured angle could account for, at most, about 18 per cent of the total values possible and measurements become more precise for increasing inclination angles. This makes sense as more extreme angles induce more observable Doppler effects on the \feka\ line. While iron abundance is a bit more difficult to constrain, it is measured within about 30 percent. Lastly, ionization and inner emissivity index are unable to be constrained in the 2 -- 10\keV, $R$ = 1 scenario. This is not unexpected due to the limited bandwidth, which essentially forces the entire reflection component to be modelled based on the \feka\ line alone. 

Extending the fits up to 70\keV, constraints do improve significantly for most reflection parameters. Photon index, iron abundance, and inclination are no longer over-estimated. Ionization also becomes reasonably constrained for values $<$200 \ergcmps. However, fit models continue to be insensitive to emissivity index.

\begin{figure*}
	\centering
	\vspace{-1in}
	\advance\leftskip-2.5in   
	\begin{minipage}{0.4\linewidth}
		\scalebox{0.35}{\includegraphics{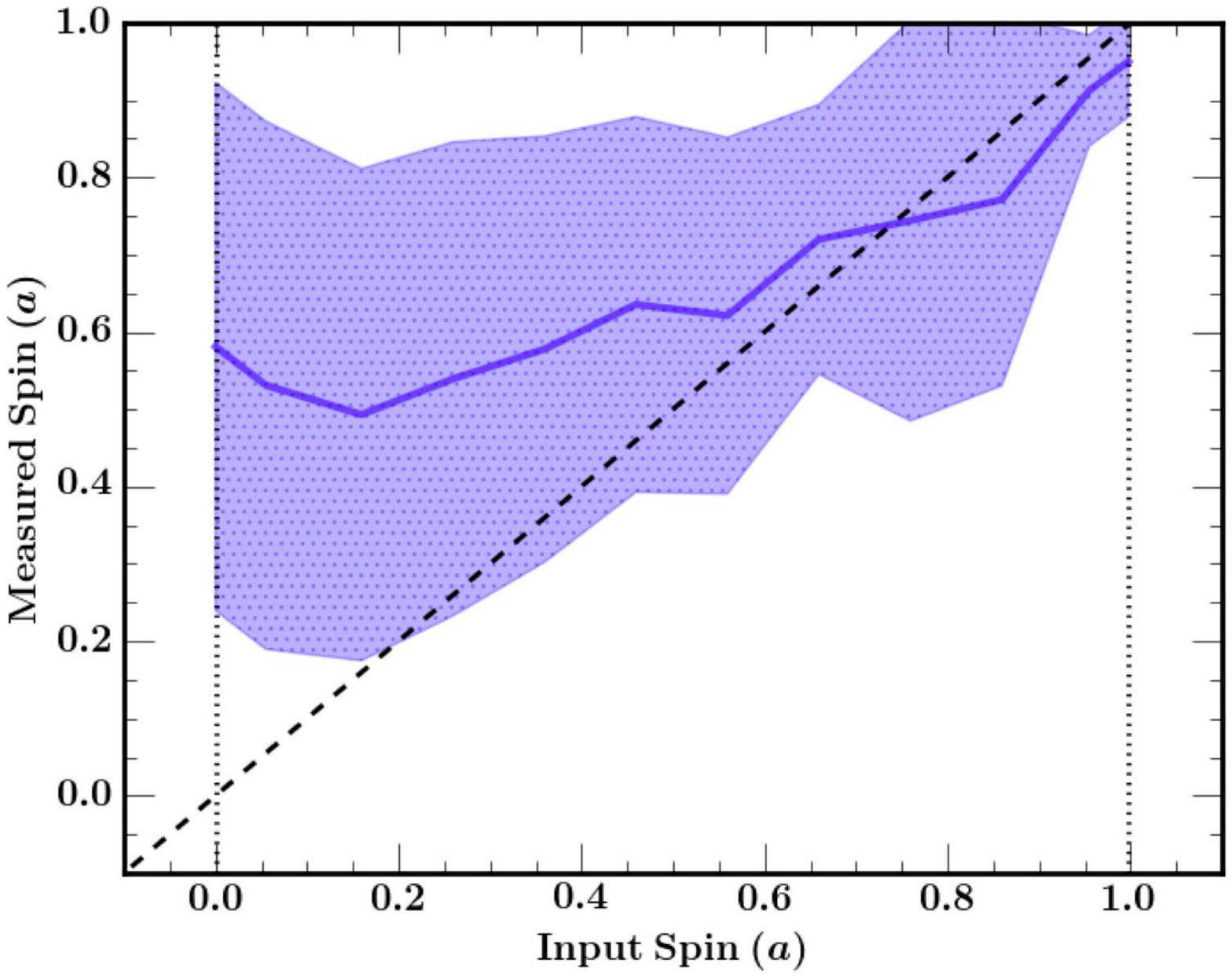}}
	\end{minipage}  
	\begin{minipage}{0.01\linewidth}
		\scalebox{0.35}{\includegraphics{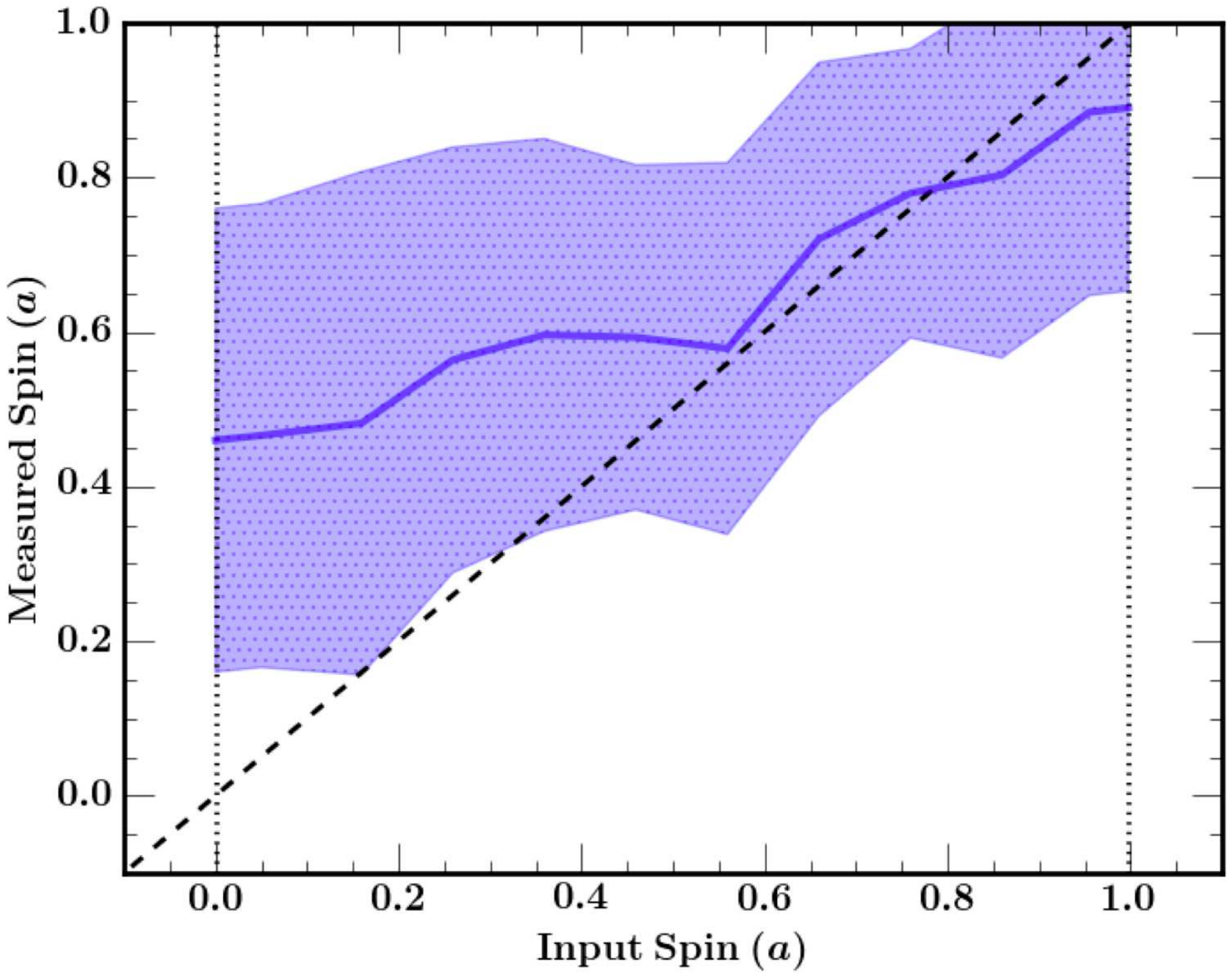}}
	\end{minipage}\\ 
	\vspace{-1.75in}
	\begin{minipage}{0.4\linewidth}
		\scalebox{0.35}{\includegraphics{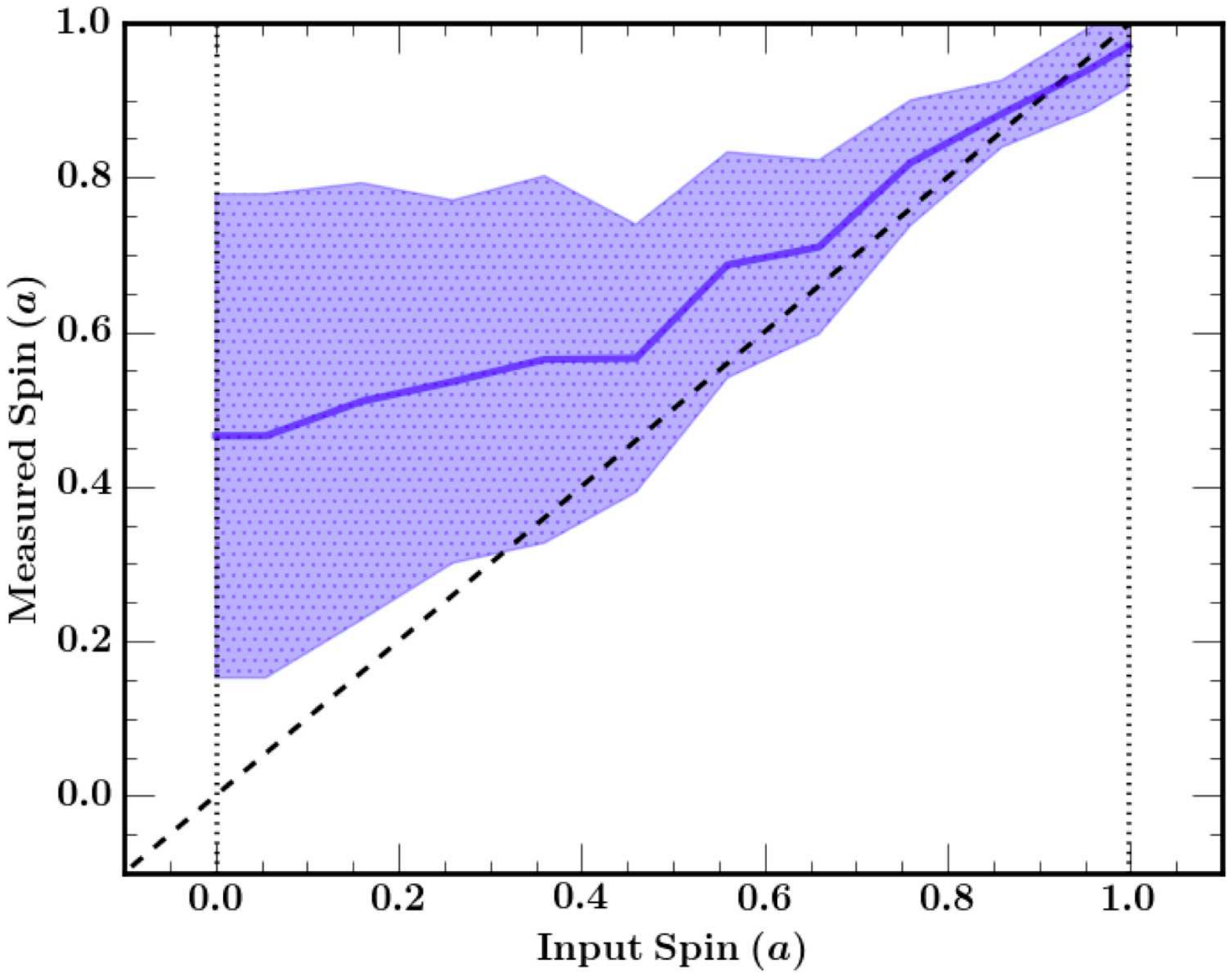}}
	\end{minipage} 
	\begin{minipage}{0.01\linewidth}
		\scalebox{0.35}{\includegraphics{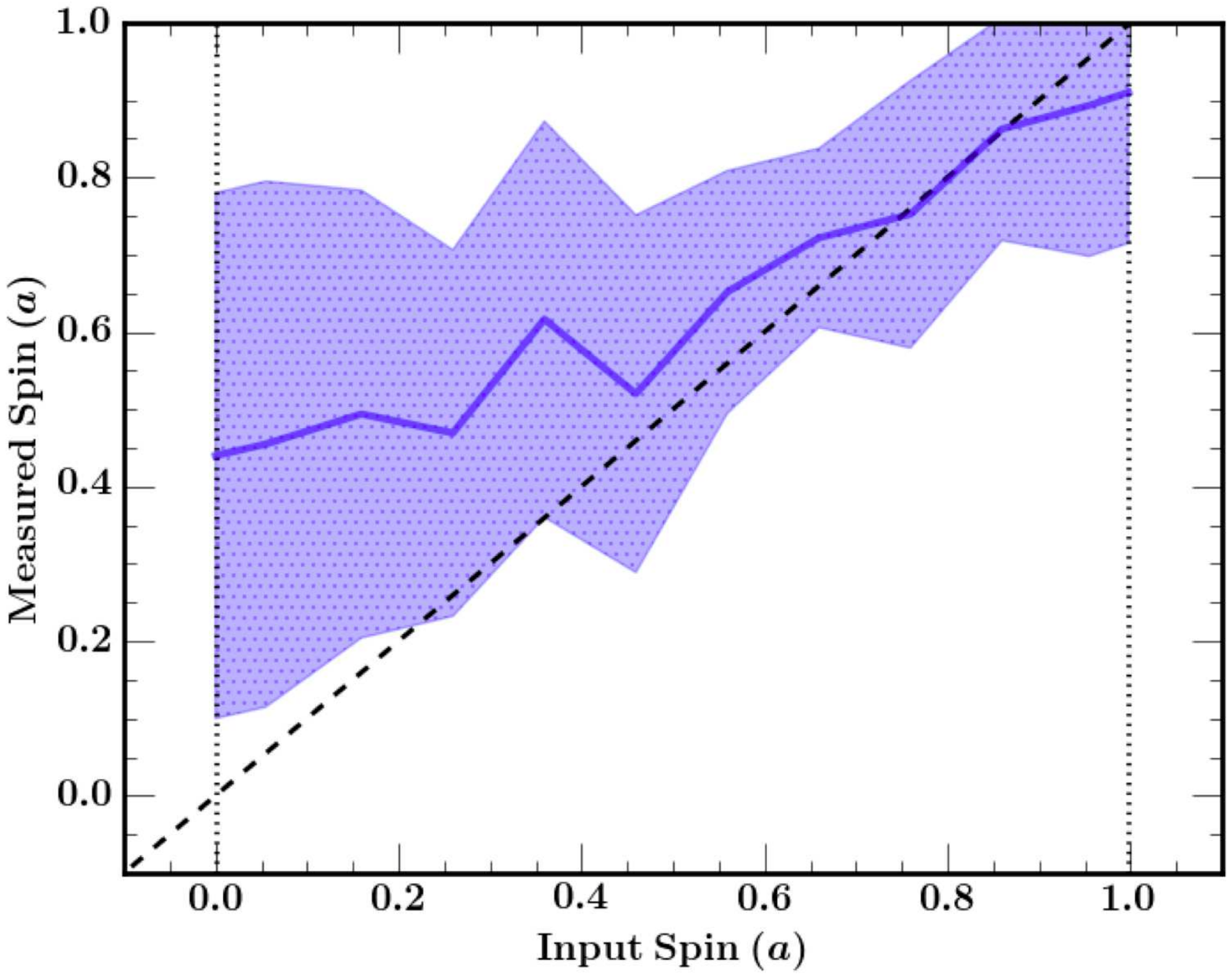}}
	\end{minipage}
	\vspace{-0.75in}
    	\caption{Summary of spin measurement results for Test A, placed side-by-side for visual comparison. Top Left: $R$ = 1 spectra fit from 2.5 -- 10.0\keV. Top Right: $R$ = 1 spectra fit from 2.5 -- 70\keV. Bottom Left: $R$ = 5 spectra fit from 2.5 -- 10\keV. Bottom Right: $R$ = 5 spectra fit from 2.5 -- 70\keV. Only prograde spins are allowed in these models. Each $R$-test uses the same simulated source spectra; fit conditions are modified while input parameters remain unchanged.}
   	 \label{SpinSum}
\end{figure*}

Comparing only the spin results from the $R$ = 1 analysis, we can begin to draw tentative conclusions about the robustness of an average AGN spin measurement. Since there is no significant improvement in parameter constraints with fit test, we continue by looking only at fit Test A, where all key parameters are left free to vary: when considering only the standard 2.5 -- 10\keV\ energy band, spin is poorly constrained and grossly over-estimated below $a$ $\sim$ 0.6 (Fig.\thinspace\ref{SpinSum}, top left). As mentioned above, additional reflection parameters such as ionization and inner emissivity index are unable to be constrained without additional information.

Extending spectral fits into the hard band up to 70\keV\ (Fig. \ref{SpinSum}, top right) provides only a small improvement to spin measurements below $a$ $\sim$ 0.6, however the random error is still substantial. It appears that the most extreme spin measurements, say $a$ $>$ 0.9, may be considered sound as the range of possible ``input'' values (i.e. the intrinsic spin) is reasonably narrow: about 30 per cent the total range of measurement values possible.

Once the lower limit for possible model $a$ values is extended and retrograde spin is allowed for in the fitting process (i.e. -0.998 $\leq$ $a$ $\leq $0.998), our ability to constrain even the most extreme spins worsens. Due to the increased random error, it appears allowing for retrograde spin diminishes our ability to measure even extreme values. It is expected that opening the lower limit for retrograde spin would increase standard deviation for smaller spin values, but doing so also had the unexpected consequence of increasing the standard deviation for larger spin values as well. The reason behind this is not immediately clear and it may be an artifact of the fitting process. For example, the larger parameter space is subject to more local minima. Therefore, including the possibility of retrograde measurements complicates the model fitting process, especially for objects with more extreme spin values --- both high and low. 

%%%%%%%%%% Figure 6: R1 Spin Summary %%%%%%%%%
%%%%%%%%%%%%%%%%%%%%%%%%%%%%%%%%%%

\subsection{$\bm{R}$5: band comparison}
\label{disc_R5}
When the 2.5 -- 10\keV\ $R$ = 5 spectra are refit up to 70\keV, we increase our ability to constrain reflection parameters $A_{Fe}$, $\theta$, and $\xi$, consistent with the results of $R$ = 1 investigation. It is interesting to note that our ability to constrain $\xi$ below values of $\sim$200\ergcmps\ greatly improves once the fit band has been extended. We can credit our improved ability to measure $\xi$ to the higher reflection fraction and, as in the $R$ = 1 scenario, that ability continues to improve with increased bandwidth. However, we do not improve our ability to constrain black hole spin. For this case of a higher reflection fraction, spin measurement precision decreases, with input range growing from $a_{in}$$\sim$0.1 to $a_{in}$$\sim$0.25 for a measured value of $a$ = 0.95.

In the case of reflection-dominated AGN, it appears that there is an advantage to measuring black hole spin in the narrower 2.5 -- 10\keV\ band (Fig.\thinspace\ref{SpinSum}, bottom left) rather than in full 2.5 -- 70\keV\  band (Fig.\thinspace\ref{SpinSum}, bottom right). This result may seem counterintuitive, however one must keep in mind that Fig.\thinspace\ref{R52N_bands} shows all other reflection component parameters are measured well with increasing bandpass range (with the exception of $q_{1}$, which is never constrained). It is reasonable to expect that reflection parameters such as $\theta$ or $\xi$ become easier to model as more of the reflection component is ``observed'' via the Compton hump. Parameters like $a$ and $A_{Fe}$ rely on spectral features in the \feka\ band and would be improved with higher signal-to-noise in the 2.5 -- 10\keV\ band in addition to broadening the bandwidth. Having a broader bandpass when measuring spin exclusively seems to confuse the modelling of the \feka\ profile and, unless spectral resolution is also increased with the bandwidth, these results suggest the standard 2 -- 10\keV\ band should be used when constraining spin in this manner; i.e. with single instrument, single epoch observations. In other words, we can be confident in our high ($a$ $>$ 0.8) spin measurements to about $\pm0.1$. It should be noted that a simultaneous \nustar and \xmm observation increases the 2 -- 10\keV\ signal-to-noise as well as extends the observable energy range.

Allowing for retrograde spin measurements in the $R$ = 5 scenario only served to worsen spin constraints when the fit band was restricted to the 2.5 -- 10\keV\ regime. While observing a real AGN with true retrograde spin remains a possibility, their assumed rarity in nature reassures us that this complication in the modelling process is reserved for special circumstances and does not affect the majority of spectroscopic analyses. 

%------------------------------------------------------------------------------------

\subsection{Caveats} 
\label{caveats}
These simulated spectral fits have elucidated the need for observers to be cautious when attempting to measure reflection parameters, especially black hole spin, via \feka\ line-fitting. While there is clear justification for similar investigations in the future, one does need to keep in mind the limitations of the work presented here. 

As stated in Section \ref{sim_details}, these simulations imitate the most ``ideal'' AGN spectrum with regards to detecting the reflection component of the X-ray spectrum (i.e. no absorption, high count rate, local object, moderate and high reflection fractions). We also fully sample the parameter space when creating the simulated spectrum; this ensures reliable error statistics, however a consequence of randomly selecting the parameter values is that we risk creating spectra with less than physical combinations of parameters (e.g. low $q_{1}$ with high $a$ and low $\xi$). An examination of parameter vs. parameter space was performed for all six key parameters investigated here to see if any unphysical or extreme combinations influenced the fitting process. No such correlation was found. It must also be kept in mind that the simulations produced here are strictly mathematical models and while they mimic average Seyfert 1 X-ray spectra, they are not intended to be substitutes for empirical data.

In addition to potentially unphysical parameter combinations, we rely exclusively on the sampling statistics as a representation of the random error in the model fits. This is a reasonable first-order assumption, however there is a risk of our \redchisq\ goodness-of-fit being misrepresented by local minima. Indeed, the greater range with which the fit could fall into local minima could easily explain the observed retrograde spin results.

In an effort to understand such effects, we investigated the role of local minima by refitting the simulated spectra from the $R$ = 1, Test A, 2.5 -- 70\keV\ scenario (Fig. \ref{R1XMM2N_bands}, blue). The spectra were refit with the same model as before however this time we included error checks on all six key parameters. While this cannot guarantee absolute minima, it reduces the likelihood of local analogues. The results of both error tests were consistent and there were no significant differences between the measurement profiles using error checks and those that do not. This does not imply that error checks in spectral model fitting are superfluous: the overall fraction of good (i.e. \redchisq $<$ 1.1) fits increased with fits that included error checks compared to those without. However, using the scatter in a larger number of spectra seem comparable to measuring errors on each parameter. Thus we did not run parameter error checks for the other tests in the interest of time.

Perhaps most importantly, AGN astronomers also do not rely exclusively on \feka\ line-fitting when performing true empirical analysis, but rather use a multi-pronged approach that often includes multi-epoch observations, timing analysis like fractional variability and reverberation mapping, and/or more robust statistical methods such as principle component analysis. In this simulated study, we have focused solely on fitting a single-epoch of spectral data within two bandpasses. If we were to measure a SMBH spin to be in any region shown by these plots to be less constrained, it is possible that a more confident estimate could still be obtained by better defining the reflection component using alternative methods. 

This work confirms that we can be most confident in our SMBH spin measurements for high values of spin, above $a$ $>$ 0.8 and it can be constrained to within $\sim$10$\%$ under the simulated conditions. It is important to note that most measurements of $a$ are, in fact, high and thus likely accurate. Most AGN that have undergone spin analysis are narrow-line Seyfert 1s that literature has shown are suspected of being reflection dominated (i.e. high $R$-value) and maximally spinning. Since brighter AGN with high spin are now shown to be easier to measure, there may be a sampling bias in AGN spin measurements and it might be difficult to determine the true spin population distribution (Vasudevan et al. 2015).
 
%------------------------------------------------------------------------------------
 
\section{Conclusions \& Future Work} 
\label{conclusion}
In summary, analysis of ``average'' AGN spectral fitting under a blurred reflection scenario has shown that accurately measuring standard X-ray spectral parameters can indeed be a challenge. If restricted to the oft-utilized \feka\ line region of 2 -- 10\keV, most parameters are over-estimated and spin itself is unconstrained for all but the most extreme values. Once the bandpass is extended up to 70\keV\ the measurements improve for most parameters, those like $\Gamma$, $A_{Fe}$, and $\theta$ are no longer over-estimated, and spin is better constrained for the highest \textbf{values}. An increase in reflection fraction improves measurements further for most reflection parameters, while decreasing our ability to constrain $\Gamma$ slightly --- as to be expected in a reflection-dominant scenario. The inner emissivity index ($q_{1}$) is never constrained under the conditions tested and likely requires detailed fitting of the \feka\ profile in order to be properly estimated. Including the soft-excess in these analyses is an interesting, but lengthy challenge and will be considered in future work.

The results discussed above are found under particular conditions. That being said, those conditions are conservative and do well at representing the standard model-fitting practice of a bright AGN source. Therefore, the fact that we seem to be less able to constrain spin for intermediate values warrants caution when making empirical spin estimates and fully justifies further investigation into AGN spectral modelling as a whole. However, it must be emphasized that black hole spin can be measured with confidence for $a > 0.8$ to about $\pm$0.1, most especially for objects with a higher reflection fraction. 

The usefulness of observatories like \nustar and the soon-to-be-launched \astroh (Takahashi et al. 2014) for AGN spectroscopy cannot be overstated. The effects of increased bandwidth, improved signal-to-noise, and high spectral resolution have not been tested here. However, we expect these to improve our ability to model AGN spectra.

% --------------------------------------------------------------------------
% --------------------------------------------------------------------------

\section*{Acknowledgments}
The authors would like to thank the referee, Dr. Chris Reynolds, for constructive comments as well as Dr. Dan Wilkins, Dr. Herman Marshall, and Dr. Dom Walton for their insightful commentary regarding this study. The \xmm\ project is an ESA Science Mission with instruments and contributions directly funded by ESA Member States and the USA (NASA).

\bibliographystyle{mn2e}

%\pagebreak

%\appendix
%\section{}
%\label{app:cal}

%\clearpage

\bsp
\label{lastpage}
\end{document}